\documentclass[fleqn,usenatbib]{mnras}

\usepackage{newtxtext,newtxmath}

\usepackage[T1]{fontenc}

\DeclareRobustCommand{\VAN}[3]{#2}
\let\VANthebibliography\thebibliography
\def\thebibliography{\DeclareRobustCommand{\VAN}[3]{##3}\VANthebibliography}

\usepackage{graphicx}	
\usepackage{amsmath}	
\usepackage{mathtools}  
\usepackage{CJK}        
\usepackage{xcolor}     

\newcommand{\appendixRef}[1]{\hyperref[#1]{Appendix~\ref{#1}}}

\newcommand{\lz}[1]{#1}

\title[Photon Bubbles in Neutron Star Atmospheres]{\lz{Radiative MHD Simulations of Photon Bubbles in} Radiation-Supported Magnetized Atmospheres of Neutron Stars \lz{with Isotropic Thomson Scattering}}

\author[L. Zhang et al.]{
Lizhong Zhang (张力中),$^{1}$\thanks{E-mail: lizhong@physics.ucsb.edu}
Omer Blaes,$^{1}$
Yan-Fei Jiang (姜燕飞)$^{2}$
\\
$^{1}$Department of Physics, University of California, Santa Barbara, CA 93106, USA\\
$^{2}$Center for Computational Astrophysics, Flatiron Institute, New York, NY 10010, USA\\
}

\date{Accepted XXX. Received YYY; in original form ZZZ}

\pubyear{2021}

\begin{document}
\begin{CJK*}{UTF8}{gbsn}

\label{firstpage}
\pagerange{\pageref{firstpage}--\pageref{lastpage}}
\maketitle

\begin{abstract}
A major uncertainty in the structure and dynamics of magnetized, radiation pressure dominated neutron star accretion columns in X-ray pulsars and pulsating ultraluminous X-ray sources is that they are thought to be subject to the photon bubble instability.  We present the results of two dimensional radiation relativistic magnetohydrodynamic simulations of a non-accreting, static atmosphere to study the development of this instability \lz{assuming isotropic Thomson scattering} in the slow diffusion regime that is relevant to neutron star accretion columns. 
\lz{Photon bubbles generally grow faster toward shorter wavelengths, until a maximum growth rate is achieved at the radiation viscosity length scale, which is generally quite small and requires high numerical resolution to simulate.}
We confirm the consistency between our simulation results and linear theory in detail, and show that the nonlinear evolution inevitably leads to collapse of the atmosphere with the higher resolution simulation collapsing faster due to the presence of shorter length scale nonlinear structures.  At least in static atmospheres with horizontally periodic boundary conditions, this resolution dependence may make simulations of the nonlinear dynamics of photon bubble instability in neutron star accretion columns challenging.  It remains to be seen whether these difficulties will persist upon inclusion of an accretion flow through the top and magnetically-confined horizontal boundaries through which photons can escape.  Our results here provide a foundation for such future work.
\end{abstract}




\section{Introduction}

Accreting X-ray pulsars consist of highly magnetized neutron stars accreting material at high rates from binary companion stars.  This material is guided by the strong magnetic field of the neutron star toward the magnetic poles, and misalignment between these poles and the neutron star spin axis results in the observed pulsations
(see, e.g. \citealt{CAB12} for recent review).  At low accretion rates, material free-falls onto the neutron star, thermalizing its kinetic energy into hot spots on the stellar surface.  At higher accretion rates, however, outward photon pressure decelerates the incoming flow, resulting in a radiation shock above the stellar surface.  Below the shock, the material forms an optically thick, subsonic settling
solution in approximate hydrostatic equilibrium between outward radiation pressure and gravity \citep{Inoue1975,1976MNRAS.175..395B}. Sideways emission from such columns can be substantially super-Eddington, but this is overcome largely by tension in the confining magnetic field \citep{Inoue1975,1976MNRAS.175..395B}, in addition to the fact that
the electron scattering opacity in a strong magnetic field can be substantially
reduced below Thomson \citep{EKS15,MUS15}.  The recently discovered pulsating ultra-luminous X-ray sources \citep{2014Natur.514..202B,2016ApJ...831L..14F,2017MNRAS.466L..48I,2017Sci...355..817I,2018MNRAS.476L..45C,2019MNRAS.488L..35S,2020ApJ...895...60R} are the most extreme examples of the ability of accreting neutron stars to radiate at substantially super-Eddington luminosities, modulo \lz{possible} beaming corrections \lz{(e.g. \citealt{King2009} and \citealt{KingLasota2019}, but see \citealt{MUS21})}.  \lz{In the most extreme cases the observationally inferred energetics may even be underestimated due to neutrino emission \citep{MUS18}.}  \lz{Transient accreting neutron stars can also for a time reach substantially super-Eddington luminosities
(e.g. \citealt{TSY17,TSY18,DOR20}).}

All current models of the settling flows of neutron star accretion columns assume a density distribution that smoothly varies in space and is stationary (e.g. \citealt{WangFrank1981,BEC07,MUS15,WES17a,WES17b,GOR21}).  However, it has been known for some time that these accretion columns are dynamically unstable to the so-called photon bubble instability \citet{1992ApJ...388..561A}, which is expected to produce strong, time-dependent density fluctuations within the column.  \lz{The presence of these fluctuations may alter the time-averaged structure of the column.  Because photons will tend to preferentially escape along low density channels, photon bubbles may also ultimately determine the angular distribution of photon emission, reduce the efficacy of radiation pressure support against gravity, and affect estimates of the accretion rates above which photons can be trapped by advection.}

\lz{A} linear analysis of \lz{the growth of infinitesimal fluctuations due to} the photon bubble instability was first explored by \citet{1992ApJ...388..561A} and \citet{1998MNRAS.297..929G}, and then extended to \lz{shorter length scales} by \citet{2003ApJ...596..509B}.  The instability \lz{grows exponentially in time at a rate that is faster for shorter length scales, presumably until the fluctuations are large enough that nonlinear effects come into play.  However, at the shortest length scales,} radiative diffusion eventually becomes fast enough to smooth out fluctuations in radiation pressure (the rapid diffusion regime).  There the growth rate levels off at \lz{length scales} of
order the gas pressure scale height, and the instability becomes a radiatively amplified slow \lz{magnetosonic} mode in the gas alone.  Even before this linear behavior was understood, \citet{2001ApJ...551..897B} analytically predicted that the nonlinear development of the instability in the rapid diffusion regime would result in trains of shock waves, and this was confirmed in detail by radiation MHD simulations of \citet{2005ApJ...624..267T}.  \lz{The rapid diffusion limit of the instability can even exist in regimes where gas pressure is comparable to radiation pressure or magnetic pressure \citep{2003ApJ...596..509B}, and its nonlinear outcome has been simulated by \citet{FER13}.}

Deep inside a neutron star accretion column, the optical depths are so high that the rapid diffusion regime is only achieved on wavelengths much smaller than the scale height of the column.  It is here that the slow diffusion regime version of the photon bubble instability first studied by \citet{1992ApJ...388..561A} is most relevant, and where the instability takes on the character of a diffusion entropy mode.
Pioneering numerical simulations of the development of this instability were first conducted by \citet{1989ESASP.296...89K}, \citet{KLE96} and \citet{1997ApJ...478..663H}.  These simulations were 2D, and assumed 1D gas motion along prescribed rigid magnetic field lines.  Radiation transport was treated within the flux-limited diffusion approximation, and accounted for nonzero chemical potential effects in a Bose-Einstein spectrum.
The linear growth of the instability agreed well with the \citet{1992ApJ...388..561A} dispersion relation and photon bubbles were found to transport energy efficiently.  \lz{\citet{KLE96} and \citet{Kleinetal1996observ} found evidence for oscillatory behavior in the nonlinear development of their simulations of accretion columns, and predicted that these might be observable as ``photon bubble oscillations" at frequencies ranging from $\sim10^2-10^4$~Hz.  Evidence for quasi-periodic oscillations at those time scales in GRO~J1744-28, Sco~X-1, and Cen~X-3 have been claimed to be consistent with photon bubbles \citep{Kleinetal1996observ,JER00}.  On the other hand, some of these detections may have been due to instrumental artifacts \citep{REV15}.  Upper limits of 0.5 percent amplitude have been placed on quasi-periodic oscillations at kHz frequencies in the bright X-ray pulsar V0332+53 \citep{REV15}.}

However, \lz{these early} simulations were done with low spatial resolution and were not run for very long.  This may be problematic given that \lz{most existing linear analyses of the photon bubble instability have the linear growth rates increasing toward shorter length scales \citep{1992ApJ...388..561A,1998MNRAS.297..929G}.  \citet{2003ApJ...596..509B} suggested that this would continue until the pressure scale height in the gas alone is achieved, but this is a very small length scale in neutron star accretion columns.  The predicted time scales of variability may therefore depend critically on achieving adequate numerical resolution.}  More recent numerical studies by \citet{2020PASJ...72...15K}, following their previous numerical work \citep[][]{2016PASJ...68...83K},
applied similar numerical treatments of 1D gas motion and radiation diffusion. In their column simulation, the gas free falls from rest and forms finger-like inflow that resembles what \citet{1989ESASP.296...89K} found. However, it is still hard to confirm the existence and effects of photon bubble physics in these numerical accretion column simulations because to capture the photon bubbles in the simulation turns out to require sufficiently high resolution.  \citet{beg06} performed an analytic study of the nonlinear development of the instability in a static column in the slow diffusion regime, and suggested that it would result in the collapse of the column on time scales shorter than the radiative diffusion time of the assumed smoothly varying equilibrium column.

Resolving the photon bubble instability is essential to studying the dynamics of neutron star accretion columns. Hence we first need to establish a numerical framework to correctly simulate the photon bubble dynamics in a static \lz{non-accreting} column, which is the purpose of this paper. Following this work, we will add accretion to the column and explore the dynamics modified by the photon bubble physics, which will be presented in a separate paper. In contrast to much previous numerical work \citep{1989ESASP.296...89K,KLE96,1997ApJ...478..663H,2016PASJ...68...83K}, our simulations do not assume 1D gas motion along rigid magnetic field lines, but instead allow for full dynamics of the magnetic field in response to fluid motions.
While not critical to the simulations reported here, this is an important aspect in preparation for
more global simulations, where the gas may not be strictly confined by the magnetic field.

Compared to previous numerical work, our simulations are able to go to much higher resolution, better resolving the faster growing modes.  Rather than using moment methods like flux-limited diffusion or M1, we also exploit a radiation transport algorithm that accurately bridges the optically thick and thin regimes by computing directly the angle dependence of the radiation intensity \lz{from the frequency-integrated radiative transfer equation} \citep{2014ApJS..213....7J}.  \lz{A major advantage of this algorithm is that it automatically incorporates the effects of radiation viscosity, which cannot be computed by flux-limited diffusion or M1.  We show here that it is radiation viscosity which ultimately limits the growth rate of photon bubbles at small length scales.  Our simulations here} will lay the ground work for modern, higher resolution global simulations of the dynamics of neutron star accretion columns.

In \hyperref[sec:pbi_overview]{Section 2}, we give a brief explanation of the photon bubble physics. In \hyperref[sec:numerical_method]{Section 3}, we introduce the numerical approaches and model configurations behind the simulations. In \hyperref[sec:results]{Section 4}, we present and explain the simulation results, compare the simulation with linear theory, and study the behavior of the photon bubble simulations at different resolutions. All the quantitative derivations and some details concerning the numerical algorithms can be found in the \hyperref[appendix]{Appendices}.

\section{Overview of Photon Bubble Instability}
\label{sec:pbi_overview}
The linear analysis of the photon bubble instability in the slow diffusion regime was explored by \citet{1992ApJ...388..561A} and \citet{1998MNRAS.297..929G}, although both of these studies also included aspects of the shorter wavelength rapid diffusion regime behavior. We begin by briefly describing the physics of the slow diffusion mode.  \lz{More details can be found in \hyperref[sec:derivation_pbi]{Appendix A}, where we present a detailed derivation of our version of the instability dispersion relation that includes the effects of radiation viscosity.  We use this dispersion relation} to compare with our numerical simulations. 

Consider a static column at one of the magnetic poles of the neutron star, and assume for simplicity that its vertical extent is much smaller than the radius of the star, so that the magnetic field can be treated as purely vertical and the gravitational acceleration $g$ can be assumed constant.  For the purposes of this physics discussion, imagine the magnetic field to be so strong as to be perfectly rigid, although we will allow for the field to be dynamic in our numerical simulations below.  Because no fluid can then move horizontally, and radiation pressure completely dominates gas pressure, hydrostatic equilibrium in the column states simply that the vertical radiation flux is $F_z=cg/\kappa$.  Taking the opacity $\kappa$ to be constant\footnote{Throughout this paper we assume that the opacity is simply that of non-magnetic Thomson scattering, neglecting the angle and polarization dependence that is in fact important for neutron star accretion columns.  We intend to incorporate these effects in future work.} this then implies that the vertical radiation flux is constant.  Assuming radiative diffusion, $F_z=-(c/3\kappa\rho)dE_r/dz$ then relates the vertical gradient in radiation energy density $E_r$ to the density $\rho$.

A key reason why the column is vulnerable to instability is that such an equilibrium is completely unchanged if we add arbitrary vertical fluctuations in density, provided we maintain a vertically constant flux $F_z=cg/\kappa$ by adjusting the radiation energy density gradient to be such that $dE_r/dz\propto\rho$.  Because such fluctuations necessarily involve fluctuations in radiation entropy, this static mode is an entropy mode.  However, the finite horizontal extent of the accretion column means that such fluctuations must themselves have horizontal variations, and the resulting horizontal radiative diffusion then causes time dependence in the density and radiation energy density.  Provided this time-dependence is slow enough that the gas inertia is truly negligible, then vertical hydrostatic equilibrium would still be maintained, and of course horizontal equilibrium is maintained by the strong vertical magnetic field.

However, the gas inertia, while small, is still finite in the slow diffusion regime, and cannot in fact be neglected.  As \lz{first shown by \citet{1992ApJ...388..561A} (see also \hyperref[sec:origin_of_pbi]{Appendix A3})}, the inertia introduces a finite perturbed vertical flux, which provides extra radiation support to balance the perturbed net force. This response of the force balancing has a $90^{\circ}$ phase delay with respect to the density perturbation (\autoref{eq:phase_delay_drho}, \autoref{eq:phase_delay_dFz} and \autoref{fig:heat_flow}), which causes a small amount of radiation to flow from high-density regions to low-density regions. Although this unstable effect is small, this tendency would eventually evacuate the perturbed low-density regions by feeding in radiation and leading to an increasing amplitude of density perturbation.

In \hyperref[sec:pbi_dispersion_relation]{Appendix A2}, we derive a short-wavelength (WKB) linear dispersion relation \lz{that fully includes the effects of radiation viscosity for the first time (\autoref{eq:dispersion_blaes_norm})}.

\lz{Numerical solutions of this dispersion relation for the instability growth rate are shown} in \autoref{fig:growthRate3D}.  \lz{Smaller length scale modes generally grow faster until reaching a maximum growth rate at a length scale $l_{\mathrm{vis}}$ (\autoref{eq:lvis}) set by radiation viscosity.}  The purpose of our simulations below will be to test our numerics against the predictions of this linear dispersion relation, and to explore the nonlinear outcome of the instability. 

\section{Numerical Method}
\label{sec:numerical_method}
\subsection{Equations}
\label{sec:equations}
\lz{Accreting neutron stars in high mass X-ray binary systems have strong surface fields $\sim 10^{12-13}$~G (e.g. \citealt{Bellmetal2014,DallOsso2015}) that may in some cases even extend up to magnetar field strengths ($10^{14}$~G,\citealt{TSY16}).  Because of this, the Alfv\'en speed in Newtonian MHD} can easily exceed the speed of light in low density regions, which slows down the numerical simulation because of the CFL condition on the time step. \lz{On the other hand,} the Alfv\'en speed in relativistic MHD is intrinsically limited to the speed of light. Therefore, we couple special relativistic MHD (RMHD, \citealt{2011ApJS..193....6B}) and the radiative transfer equation \citep[][]{2014ApJS..213....7J} in the Athena++ code \citep[][]{STO20}, and solve
them together for the radiation pressure dominated static column on the neutron star.  We provide details of our modifications to the Athena++ algorithms in \hyperref[sec:code_modify]{Appendix C}. 

The primitive variables $(\rho, v^i, P_g, B^i)$ in RMHD are defined in the fluid rest frame, where $\rho$ is the gas density, $P_g$ is the gas pressure, $v^i$ is the fluid three-velocity, and $B^i$ is the magnetic field three-vector. Hereafter, we use Latin indices in italics to denote spatial components of three-vectors (from 1 to 3) and Greek indices to denote components of four-vectors (from 0 to 3), where 0 represents the time component. We adopt velocity units with $c=1$ and Minkowski metric $\eta_{\mu\nu}=\mathrm{diag}(-1,1,1,1)$ for flat spacetime. Given a Lorentz factor defined as $\Gamma = (1-v_jv^j)^{-1/2}$, the fluid four-velocity components are $u_0=\Gamma$ and $u^i = \Gamma v^i$. The magnetic field four-vector $b^{\mu}$ and the total enthalpy $w$ are defined as follows for convenience
\begin{align}
    b^0 &= u_jB^j,\quad b^i = \frac{1}{\Gamma}(B^i + b^0u^i)
    \quad, 
    \\
    w &= \rho +\frac{\gamma}{\gamma-1}P_g +b_{\nu}b^{\nu}
    \quad, 
    \label{eq:total_enthalpy}
\end{align}
where the equation of state for ideal gas is adopted in (\autoref{eq:total_enthalpy}) with the gas adiabatic index $\gamma$. 
The system is governed by the gas conservation laws and the radiative transfer equation. We summarize these equations below in the sequence of particle number conservation, momentum conservation, energy conservation and radiative transfer.
\begin{align}
    &\partial_0(\rho u^0) + \partial_j(\rho u^j) = S_{\mathrm{gr}1}
    \quad, 
	\label{eq:particle_conserv}
	\\
	&\begin{multlined}[t]
	\partial_0(w u^0u^i - b^0b^i)
	\\
	+ \partial_j\left(w u^iu^j + \left(P_g+ \frac{1}{2}b_{\nu}b^{\nu}\right)\delta^{ij} - b^ib^j \right) = S_{\mathrm{gr}2}^i - S_{r2}^i
    \quad, 
	\end{multlined}
	\label{eq:mom_conserv}
	\\
	&\begin{multlined}[t]
	\partial_0\left[w u^0u^0 - \left(P_g+ \frac{1}{2}b_{\nu}b^{\nu}\right) - b^0b^0\right]
	\\
	\mkern155mu + \partial_j(w u^0u^j - b^0b^j) = S_{\mathrm{gr}3} - S_{r3}
    \quad, 
	\end{multlined}
	\label{eq:energy_conserv}
	\\
	&\partial_0I + n^j\partial_j I = \mathcal{L}^{-1}(\bar{S}_r)
    \quad, 
	\label{eq:rad_transfer}
\end{align}
where $S_{\mathrm{gr}1}$, $S_{\mathrm{gr}2}^i$ and $S_{\mathrm{gr}3}$ are the gravitational source terms that mock up the gravity in special relativity. $I$ is the frequency-integrated intensity and the unit vector $n^i$ is the direction of the intensity. $\mathcal{L}$ is the Lorentz boost operator from the lab frame to the fluid frame and $\mathcal{L}^{-1}$ vice versa. $S_{r2}^i$ and $S_{r3}$ are the momentum and energy exchange between gas and radiation. The radiative transport term $\bar{S}_r$ is defined in the fluid frame including the processes of elastic scattering, absorption and Compton scattering. Note that the coordinates and radiation variables are defined in the lab frame and we denote these quantities in the fluid frame with overbars. 

When the plasma is radiation pressure dominated and the Lorentz factor is near unity or varies slowly in spacetime, we can treat the gravitational field near the neutron star surface by using the following approximate source terms in RMHD (see derivations in \hyperref[sec:grav_source_terms]{Appendix B}). 
\begin{align}
    S_{\mathrm{gr}1} =& (2\Gamma^2+1)\rho u^j\partial_j\phi
    \quad,
    \\
    S_{\mathrm{gr}2}^i =& 2(2\Gamma^2+1)w_gu^iu^j\partial_j\phi - (2\Gamma^2-1)w_g\partial_i\phi
    \quad,
    \\
    S_{\mathrm{gr}3} =& 2\Gamma(2\Gamma^2-1)w_gu^j\partial_j\phi
    \quad,
\end{align}
where $\phi=-GMr^{-1}$ is the Newtonian gravitational potential and $w_g = \rho + \gamma(\gamma-1)^{-1}P_g$ is the gas enthalpy. 

The radiative transfer equation is solved in the mixed frame by operator splitting the advection and source term steps.  The source term $\bar{S}_r$ is used to update the intensity in the fluid frame. Then we Lorentz transform back to the lab frame and compute the momentum ($S_{r2}^i$) and energy exchange ($S_{r3}$) between gas and radiation to update the gas primitive variables. The momentum and energy exchange between gas and radiation are 
\begin{align}
    S_{r2}^i &= \oint \mathcal{L}^{-1}(\bar{S}_r) n^i d\Omega
    \quad, 
    \\
    S_{r3} &= \oint \mathcal{L}^{-1}(\bar{S}_r) d\Omega
    \quad, 
\end{align}
where $\Omega$ is the solid angle of the radiation field. 
The radiative transport source term in the fluid frame is 
\begin{align}
    \begin{split}
        \bar{S}_r = \Gamma(1-v_jn^j)\bigg[ &\rho\kappa_s(\bar{J}-\bar{I})
        \\
        +&\rho\kappa_R\left(\frac{a_r T_g^4}{4\pi}-\bar{I}\right) +\rho(\kappa_P-\kappa_R)\left(\frac{a_r T_g^4}{4\pi}-\bar{J}\right)
        \\
        +&\rho\kappa_s\frac{4(T_g-\bar{T}_r)}{T_e}\bar{J} \bigg]
    \quad, 
    \end{split}
\end{align}
where \lz{$\kappa_s=0.34\ \mathrm{cm^2\ g^{-1}}$ is the electron scattering opacity for a fully ionized plasma}. $\kappa_R$ and $\kappa_P$ are the Rosseland and Planck mean \lz{thermal absorption} opacities, which can be numerically computed following the approximations in \citealt{2009ApJ...691...16H}.\footnote{\lz{It turns out that our results are not sensitive to these approximations, as electron scattering dominates over the Rosseland absorption opacity, and Compton scattering dominates over the Planck mean, producing enough gas-radiation energy exchange that no significant departures from LTE occur.}} 
Other quantities are the radiation density constant $a_r$, the fluid frame zeroth angular moment of intensity $\bar{J} = (4\pi)^{-1}\oint \bar{I} d\bar{\Omega}$, and the effective temperature of the radiation $\bar{T}_r = (4\pi\bar{J}/a_r)^{1/4}$, and the electron rest mass energy expressed as a temperature
$T_e$. The gas temperature is $T_g=P_g(\rho R)^{-1}$, where $R$ is the ideal gas constant \lz{assuming solar abundance}.  Note that the factor $\Gamma(1-\boldsymbol{v}\cdot\boldsymbol{n})$ outside the square brackets comes from the frame transformation. The first term in the square brackets refers to the elastic scattering. The second and third terms represent absorption or emission process. The last term is an approximation for the gas-radiation heat exchange via Compton scattering (\citealt{2003ApJ...596..509B}; \citealt{2009ApJ...691...16H}). In \hyperref[sec:code_modify]{Appendix C}, we briefly describe how to solve radiation transport numerically and how we modify it for special relativity based on the work done by \citet{2014ApJS..213....7J}. 
Note that we assume isotropic opacities and neglect the angle and polarization dependence. We use a \lz{frequency-averaged} treatment of radiation transfer with an assumed blackbody spectrum and neglect photon chemical potential and Bose-Einstein effects.  \lz{We also neglect QED effects and electron-positron pair production, as we never achieve temperatures here where pair production will be important. This can be an important effect for high magnetic field strengths and/or higher temperatures in real accretion columns in ultraluminous X-ray sources, and will likely have a strong effect
on photon bubble dynamics simply by reducing the Eddington limit \citep{MUS19}.  On the other hand, photon bubbles might enhance radiation escape and lower the temperature inside the accretion column.}

\subsection{Simulation Domain, Initial and Boundary Conditions}

\begin{figure*}
	\includegraphics[width=\textwidth]{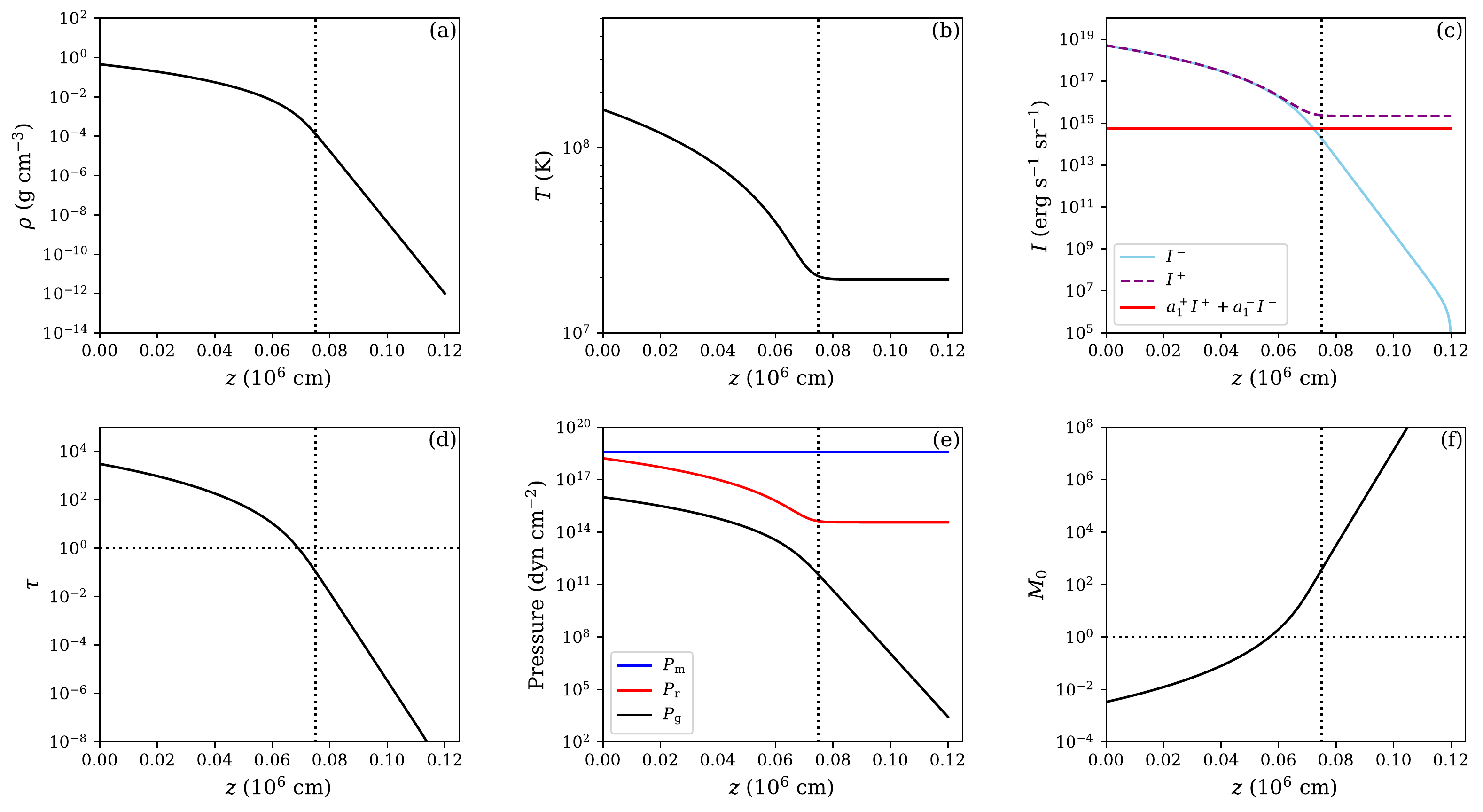}
    \caption{The \lz{vertical} profiles of \lz{various quantities in the initial condition of} the atmosphere\lz{:  (a) density; (b) gas temperature, which equals the radiation effective temperature as the initial condition is assumed to be in LTE; (c) upward ($I^+$) and downward ($I^-$) two-stream intensities along with vertical radiation flux divided by the speed of light ($F_r/c=a_1^+I^++a_1^-I^-$); (d) vertical Rosseland mean optical depth; (e) magnetic ($P_{\rm m}$), radiation ($P_{\rm r}$), and gas ($P_{\rm g}$) pressures; and (f) the radiation diffusion Mach number (see \autoref{eq:auxiliary_param}).  The vertical dotted line in all panels indicates the location where the vertical Rosseland mean optical depth $\tau=0.1$.  Horizontal dotted lines in panels (d) and (e) indicate where $\tau=1$ and $M_0=1$, respectively.}
    }
    \label{fig:ini_condition}
\end{figure*}

We adopt a Cartesian geometry in a plane-parallel atmosphere on the neutron star surface near a magnetic pole.
We initialize the simulation with a vertical magnetic field and the fluid in vertical hydrostatic equilibrium.  Although our simulation uses RMHD, we neglect this and also assume a constant Newtonian gravitational acceleration for our initial condition. \lz{For the purposes of setting up the initial condition only, we assume} local thermal equilibrium (LTE) between the gas temperature and radiation effective temperature $T_g=T_r\equiv T$, and a standard Eddington closure scheme between the zeroth and second angular moments of the radiation field.   We can then integrate the zeroth and first moments of the radiative transport equation to obtain the energy and momentum equations for the radiation. 
Meanwhile, the equations of conservation of particle number (\autoref{eq:particle_conserv}) and energy (\autoref{eq:energy_conserv}) become trivial because of $v^i=0$ and LTE.
Then hydrostatic equilibrium can be described by the following equations 
\begin{align}
    \partial_zP_g &= -\rho\left(g-\kappa_F\frac{F_r}{c}\right)
    \quad, 
    \\
    \partial_zF_r &= 0, \ E_r = a_r T^4
    \quad, 
    \\
    \partial_zP_r &= -\rho\kappa_F\frac{F_r}{c}
    \quad, 
\end{align}
where $E_r$, $F_r$ and $P_r$ are the radiation energy density, radiation flux and radiation pressure respectively. The gravitational acceleration $g=1.86\times10^{14}\ \mathrm{cm\ s^{-2}}$ is for the surface of neutron star. 
The flux mean opacity $\kappa_F=\kappa_s+\kappa_R$ is the effective opacity for gas-radiation momentum coupling. In order to close the system in hydrostatic equilibrium and obtain the smooth transition near the photosphere, we initialize the radiation field by the two-stream approximation as follows
\begin{align}
    E_r &\equiv a_0^{+} I^{+} + a_0^{-} I^{-}
    \quad, 
    \\
    F_r/c &\equiv a_1^{+} I^{+} + a_1^{-} I^{-}
    \quad, 
    \\
    P_r &\equiv a_2^{+} I^{+} + a_2^{-} I^{-}
    \quad,
\end{align}
where $I^{+}$ refers to the direction $n_z>0$ and $I^{-}$ refers to the direction $n_z<0$. Here $n_z = \cos{\theta_r}$ and $\theta_r$ is the polar angle. The coefficients $a_0^{\pm}$, $a_1^{\pm}$ and $a_2^{\pm}$ are constants and depend on the numerical setup of the discrete solid angles of the intensity field. 
\begin{align}
    a_0^{+} &= \sum w^{+}\quad, &&a_0^{-} = \sum w^{-}
    \quad, 
    \\
    a_1^{+} &= \sum w^{+} n_z^{+}\quad, &&a_1^{-} = \sum w^{-} n_z^{-}
    \quad, 
    \\
    a_2^{+} &= \sum w^{+} n_z^{+}n_z^{+}\quad, &&a_2^{-} = \sum w^{-} n_z^{-}n_z^{-}
    \quad, 
\end{align}
where $n_z^{+}$ and $n_z^{-}$ correspond to the direction with $n_z>0$ and $n_z<0$, respectively. The quantities $w^{\pm}$ are the weights corresponding to the directions $n_z^{\pm}$. Therefore, we can rewrite the hydrostatic equilibrium for the numerical purpose as follows
\begin{align}
    \partial_z\rho &= -\frac{(1-\epsilon)\rho g}{RT} - \left( \frac{a_0^{-}a_1^{+}}{a_1^{-}a_2^{+}-a_1^{+}a_2^{-}} + \frac{a_0^{+}a_1^{-}}{a_1^{+}a_2^{-}-a_1^{-}a_2^{+}} \right)\frac{\epsilon\rho g}{4a_rT^4}
    \quad, 
    \\
    \partial_zT &= \left( \frac{a_0^{-}a_1^{+}}{a_1^{-}a_2^{+}-a_1^{+}a_2^{-}} + \frac{a_0^{+}a_1^{-}}{a_1^{+}a_2^{-}-a_1^{-}a_2^{+}} \right)\frac{\epsilon\rho g}{4a_rT^3}
    \quad, 
    \\
    \partial_zI^{+} &= \left(\frac{a_1^{+}}{a_1^{-}} a_2^{-} - a_2^{+}\right)^{-1} \epsilon\rho g
    \quad, 
    \\
    \partial_zI^{-} &= \left(\frac{a_1^{-}}{a_1^{+}} a_2^{+} - a_2^{-}\right)^{-1} \epsilon\rho g
    \quad, 
\end{align}
where $\epsilon$ is the local Eddington ratio, which indicates the fraction of the gravity supported by the radiation support. Then the system can be integrated from the top given
\begin{align}
    \rho|_{z_{\infty}} &= 0
    \quad, 
    \\
    T|_{z_{\infty}} &= \left(\frac{a_0^{+}}{a_1^{+}}\frac{\epsilon g}{a_r \kappa_F}\right)^{\frac{1}{4}}
    \quad, 
    \\
    I^{+}|_{z_{\infty}} &= \frac{\epsilon g}{a_1^{+} \kappa_F}
    \quad, 
    \\
    I^{-}|_{z_{\infty}} &= 0
    \quad, 
\end{align}
where we adopt $z_{\infty}=0.12R_{\star}$ and $\epsilon=0.994$, with a neutron star radius $R_{\star}=10\ \mathrm{km}$. The initial conditions are numerically integrated as shown in \autoref{fig:ini_condition}. In order to resolve the gas pressure in the simulation, we want to keep the ratio of the gas pressure to the magnetic pressure ($P_m=b_{\nu}b^{\nu}/2$) as large as possible. Therefore, we adopt the magnetic field $B=10^{10}\ \mathrm{Gauss}$ and then select the top of the domain to be where the optical depth is 0.1, which is shown as the vertical dotted line in \autoref{fig:ini_condition}. This selection guarantees that the atmosphere would be well confined by the magnetic field from $z/R_{\star}=0.02$ to the top of the domain. Moreover, it covers a range of vertical optical depths from slow to rapid radiative diffusion.

We use periodic boundary conditions for the gas and radiation at the two sides. At the top, we use an outflow boundary condition for the gas, which zeros the gradient of primitive variables at the top. However, we forbid the gas to flow into the domain from the top ghost zones: 
\lz{
\begin{align}
    \rho_{(\mathrm{k_{\mathrm{max}}+k'})} &= \rho_{\mathrm{(k_{\mathrm{max}})}}
    \quad,
    \\
    v_{x (\mathrm{k_{\mathrm{max}}+k'})} &= v_{x (\mathrm{k_{\mathrm{max}}})}
    \quad,
    \\
    v_{z (\mathrm{k_{\mathrm{max}}+k'})} &= \begin{cases}
                            v_{z (\mathrm{k_{\mathrm{max}}})}\quad \textrm{, if } v_{z (\mathrm{k_{\mathrm{max}}})}\ge 0
                            \\
                            0 \qquad\quad\,\,\,\, \textrm{, if } v_{z (\mathrm{k_{\mathrm{max}}})}<0
                            \end{cases}
    \quad,
    \\
    P_{g (\mathrm{k_{\mathrm{max}}+k'})} &= P_{g (\mathrm{k_{\mathrm{max}}})}
    \quad,
\end{align}
where the subscript $(\mathrm{k_{\mathrm{max}}})$ refers to the index of $z$ that corresponds to the highest grid cell in the active zone and $(\mathrm{k'})$ starts from the first to the last grid cell in the ghost zone.
}
We use a vacuum boundary condition for the radiation at the top, which zeros the inward intensity at the top. This is a decent condition for the optical depths near the top, where the intensity is mostly outwards: 
\lz{
\begin{equation}
    I_{(\mathrm{k_{\mathrm{max}}+k'})} = \begin{cases}
                        I_{(\mathrm{k_{\mathrm{max}}})} \quad \textrm{, if } n_z\ge 0
                        \\
                        0 \qquad\quad\, \textrm{, if } n_z<0
    \end{cases} 
    \quad.
\end{equation}
}
At the bottom, we use a reflecting boundary condition for the gas. 
\lz{
\begin{align}
    \rho_{(\mathrm{k_{\mathrm{min}}-k'})} &= \rho_{\mathrm{(k_{\mathrm{min}}+k'-1)}}
    \quad,
    \\
    v_{x (\mathrm{k_{\mathrm{min}}-k'})} &= v_{x (\mathrm{k_{\mathrm{min}}+k'-1})}
    \quad,
    \\
    v_{z (\mathrm{k_{\mathrm{min}}-k'})} &= -v_{z (\mathrm{k_{\mathrm{min}}+k'-1})}
    \quad,
    \\
    P_{g (\mathrm{k_{\mathrm{min}}-k'})} &= P_{g (\mathrm{k_{\mathrm{min}}+k'-1})}
    \quad,
\end{align}
where the subscript $(\mathrm{k_{\mathrm{min}}})$ refers to the index of $z$ that corresponds to the lowest grid cell in the active zone. 
}
The radiation boundary condition at the bottom is determined by enforcing hydrostatic equilibrium there \lz{in the two-stream approximation}. 

\begin{table*}
	\centering
	\begin{tabular}{cccccccc}
		\hline
		Version & Name & Mesh & Width & Height & Resolution & Duration\\
		& & & $(R_{\star})$ & $(R_{\star})$ & ($\mathrm{cm^2/grid\, cell}$) & ($t_{\mathrm{sim}}$)\\
		\hline
		0 & LR & $\:\:340\times512\quad$ & 0.0500 & 0.075 & $147\times146$ & 1000\\
		1 & MR & $\:\:680\times1024\:\:$ & 0.0500 & 0.075 & $74\times73$ & 1000\\
		2 & HR & $\:\:680\times3072\:\:$ & 0.0167 & 0.075 & $25\times24$ & 300\\
		3 & SHR & $1280\times9472\:\:$ & 0.0100 & 0.075 & $8\times8$ & 200\\
		4 & UHR & $2560\times18944$ & 0.0100 & 0.075 & $4\times4$ & 200\\
		\hline
	\end{tabular}
	\caption{\lz{Domain size, resolution, and duration of all the simulations.}}
	\label{tab:sim_param}
\end{table*}

\lz{
\begin{align}
    & I_{(\mathrm{k_{\mathrm{min}}-k'})} = I_{(\mathrm{k_{\mathrm{min}}-k'+1})}
    \nonumber
    \\
    &\mkern55mu +\begin{dcases}
        \left(a_2^{+}-\frac{a_1^{+}}{a_1^{-}} a_2^{-}\right)^{-1} \epsilon\rho_{(\mathrm{k_{\mathrm{min}}-k'+1})}g \Delta z
        \textrm{, if } n_z\ge 0
        \\
        \left(a_2^{-}-\frac{a_1^{-}}{a_1^{+}} a_2^{+}\right)^{-1} \epsilon\rho_{(\mathrm{k_{\mathrm{min}}-k'+1})}g \Delta z
        \textrm{, if } n_z<0
    \end{dcases} 
    \quad.
\end{align}
}
In the initial condition, this leads to an upward constant radiation flux, which arises from the assumed constant Eddington ratio $\epsilon$. Then with the local primitive variables, we can compute the $I^{\pm}$ in the bottom ghost zones with first-order accuracy as follows
\begin{align}
    I^{+}_{(\mathrm{k}-1)} &= I^{+}_{(\mathrm{k})} + \left(a_2^{+}-\frac{a_1^{+}}{a_1^{-}} a_2^{-}\right)^{-1} \epsilon\rho_{(\mathrm{k})}g \Delta z
    \quad, 
    \\
    I^{-}_{(\mathrm{k}-1)} &= I^{-}_{(\mathrm{k})} + \left(a_2^{-}-\frac{a_1^{-}}{a_1^{+}} a_2^{+}\right)^{-1} \epsilon\rho_{(\mathrm{k})}g \Delta z
    \quad, 
\end{align}
where subscript $\mathrm{(k)}$ refers to the index of $z$ \lz{in the active zone}. 

We set up \lz{five} simulations with different \lz{domain sizes,} resolutions \lz{and durations as shown in \autoref{tab:sim_param}.} \lz{All simulations evolve toward collapse of the atmosphere, but we are only able to investigate this in detail in the nonlinear regime in the low-resolution (LR) and medium-resolution (MR) simulations, which we run for $1000 t_{\mathrm{sim}}$. Here $t_{\mathrm{sim}} = 2.8\times10^{-7}\ \mathrm{s}$ is the simulation time unit, roughly corresponding to the (vacuum) light crossing time across a radiation pressure scale height. The three higher resolution simulations provide a resolution study. Therefore, we only evolve the instability within the linear regime and run high-resolution (HR), super-high-resolution (SHR) and ultra-high-resolution (UHR) simulations for $300 t_{\mathrm{sim}}$, $200 t_{\mathrm{sim}}$ and $200 t_{\mathrm{sim}}$, respectively, given limited computational resources.} \lz{All} simulations are launched by applying cell-to-cell random perturbations on the initial profile in all variables with a fractional amplitude $10^{-3}$. 

There are two reasons that we prefer to apply random perturbations rather than excite one single mode: 1. the simulation is initialized by using the approximate hydrostatic equilibrium, which deviates from the true solution. Such deviation and the numerical error from the discretization of simulation grids would both contribute to the actual initial perturbation. So the linear instability would be dominated by a particular short-wavelength mode from this systematic perturbation. Therefore, we apply the cell-to-cell random perturbation to broaden the spectrum of initial perturbations.  2. in the neutron star column accretion problem, the perturbations are intrinsically random. So studying the instability in the random perturbation would provide more insight to understand the multi-mode behavior of in such system. 

\section{Results}
\label{sec:results}

In the following subsections, we present and evaluate the simulation results. In \hyperref[sec:results_overview]{Section 4.1}, we \lz{adopt the MR simulation as our fiducial simulation to} give an overall description and explanation of the evolution of photon bubble instability. In \hyperref[sec:sanity_check]{Section 4.2}, we use \lz{both the MR and UHR simulations to} evaluate the consistency between the simulation results and the analytical \lz{dispersion relation}. In \hyperref[sec:resolution_dep]{Section 4.3}, we compare the simulations at different resolutions and discuss the discrepancies introduced by finite resolution. \lz{Note that the resolution dependence is studied in the linear regime with all five simulations but in the nonlinear regime with only the LR and MR simulations.}

The photon bubble instabilities appear in \lz{all five} simulations. Animations are available online\footnote{\url{https://youtube.com/playlist?list=PLbQOoEY0CFpW276rfp1Wuzc0uoE6726cO}}. Note that in \hyperref[sec:results_overview]{Section 4.1} and \hyperref[sec:sanity_check]{Section 4.2}, we only focus on the \lz{MR} simulation and use it to study the photon bubble instability for the following reasons: 1. only \lz{the LR} and \lz{MR} simulations \lz{are evolved long enough for the atmosphere to collapse} due to the photon bubble instability. 2. \lz{compared with the LR simulation,} the \lz{MR} simulation is less noisy and has better numerical performance. 3. the \lz{MR} simulation can resolve shorter wavelengths than \lz{the LR simulation}, where the photon bubble instability grows faster.

\subsection{Overview of Evolution}
\label{sec:results_overview}

\begin{figure}
	\includegraphics[width=\columnwidth]{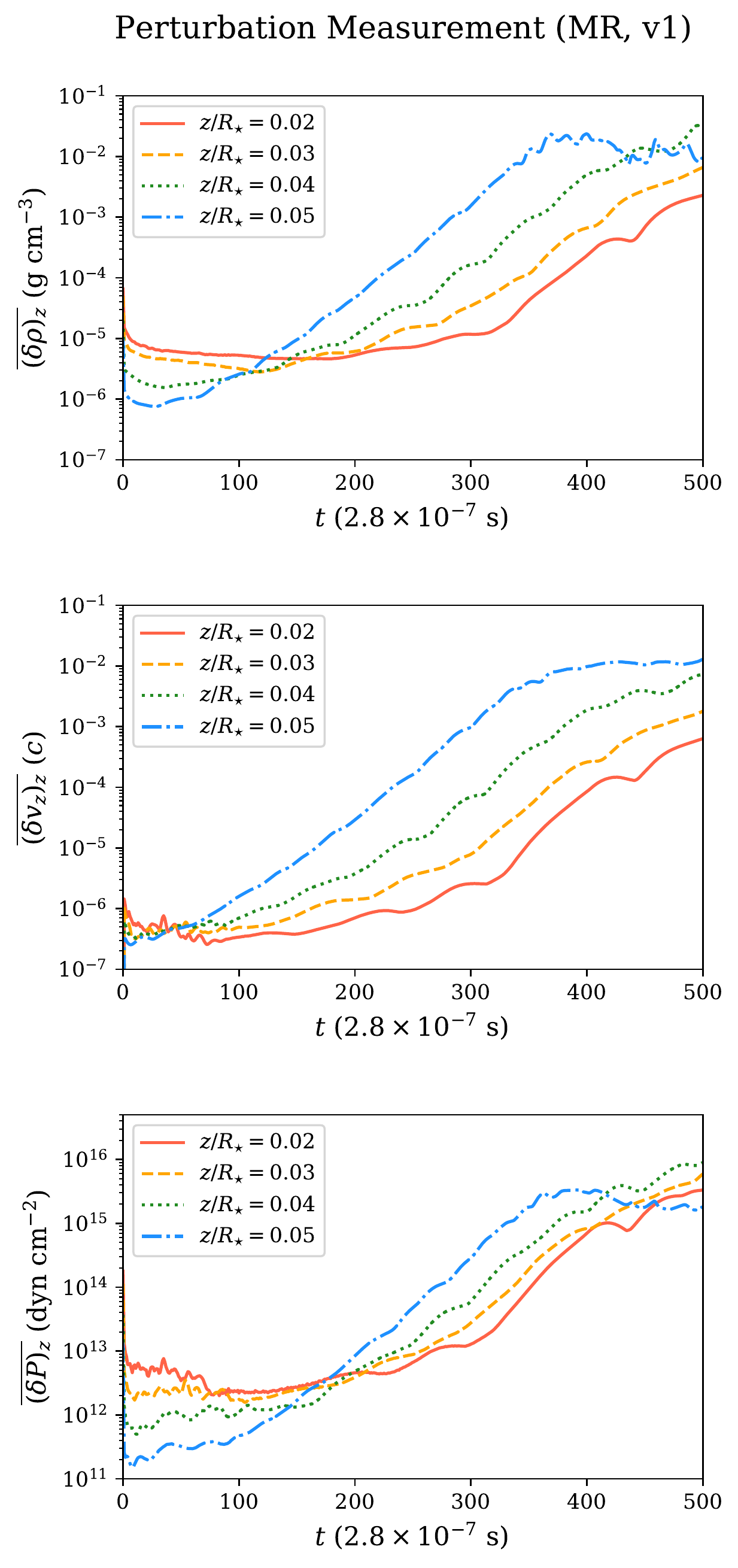}
    \caption{
    Horizontal average of the absolute value of perturbations with respect to the unperturbed initial profile in density (top), vertical velocity (middle) and thermal pressure (bottom) as a function of time \lz{and for various altitudes in the simulation as indicated.}
    }
    \label{fig:growth_rate}
\end{figure}

We track the fluctuations in terms of characteristic variables at each snapshot to study the evolution of the photon bubble instability. In the simulation, the instability patterns gradually appear from the top to the bottom because the instability growth rate is different at each height. The instability grows faster and first becomes nonlinear towards the top, since the radiation diffuses more rapidly as the gas density decreases. As the instability grows, gas and radiation are gradually spatially decoupled. Such decoupling eventually makes the gas lose the radiation support and sink down due to the gravity. Then the whole atmosphere collapses and the radiation freely leaves the domain at the top. 

Before the nonlinear phase of the instability starts to dominate the system, we select four different heights ($z/R_{\star}=0.02$, $0.03$, $0.04$ and $0.05$, where the magnetic field continues to be strong enough to constrain the gas to move vertically) to study the linear growth of the instability by monitoring the horizontally averaged perturbations in density, velocity and pressure. Specifically, we use $v_z$ and $P$ to denote the velocity in $z$-direction and the total pressure of the gas and radiation respectively. Note that $P\simeq P_r$ because the regime is radiation-dominated. As shown in the \autoref{fig:growth_rate}, in the beginning, the initial perturbations at each height are relaxed by the system since the initial condition is not in the perfect hydrostatic equilibrium. After the system is relaxed close to the hydrostatic equilibrium, the linear instability starts to grow independently at each height, growing fastest in the highest altitude regions. As the linear instability grows, the fluctuations at different heights propagate and interfere with each other. During such interference, one region would be dominated by another. In our case, fluctuations in higher regions always dominate lower regions because of the higher instability growth rate. Therefore, we can approximately identify when the interference happens by tracing the time that growth rates at two adjacent heights start to synchronize (\autoref{tab:t_syn}), where we denote such time as $t_{\mathrm{syn}}$. As the simulation continues, the photon bubble instability becomes nonlinear and thus we stop tracking the perturbations.

\begin{table}
	\centering
	\begin{tabular}{lccc}
		\hline
		$z/R_{\star}$ & $0.05\rightarrow 0.04$ & $0.04\rightarrow 0.03$ & $0.03\rightarrow 0.02$\\
		\hline
		$t_{\mathrm{syn}}/t_{\mathrm{sim}}$ & $\sim 310$ & $\sim 330$ & $\sim 350$\\
	\end{tabular}
	\caption{The approximate times when the linear growth rates synchronize between two adjacent heights in the \lz{MR} simulation. }
	\label{tab:t_syn}
\end{table}

\begin{figure*}
	\includegraphics[width=0.82\textwidth]{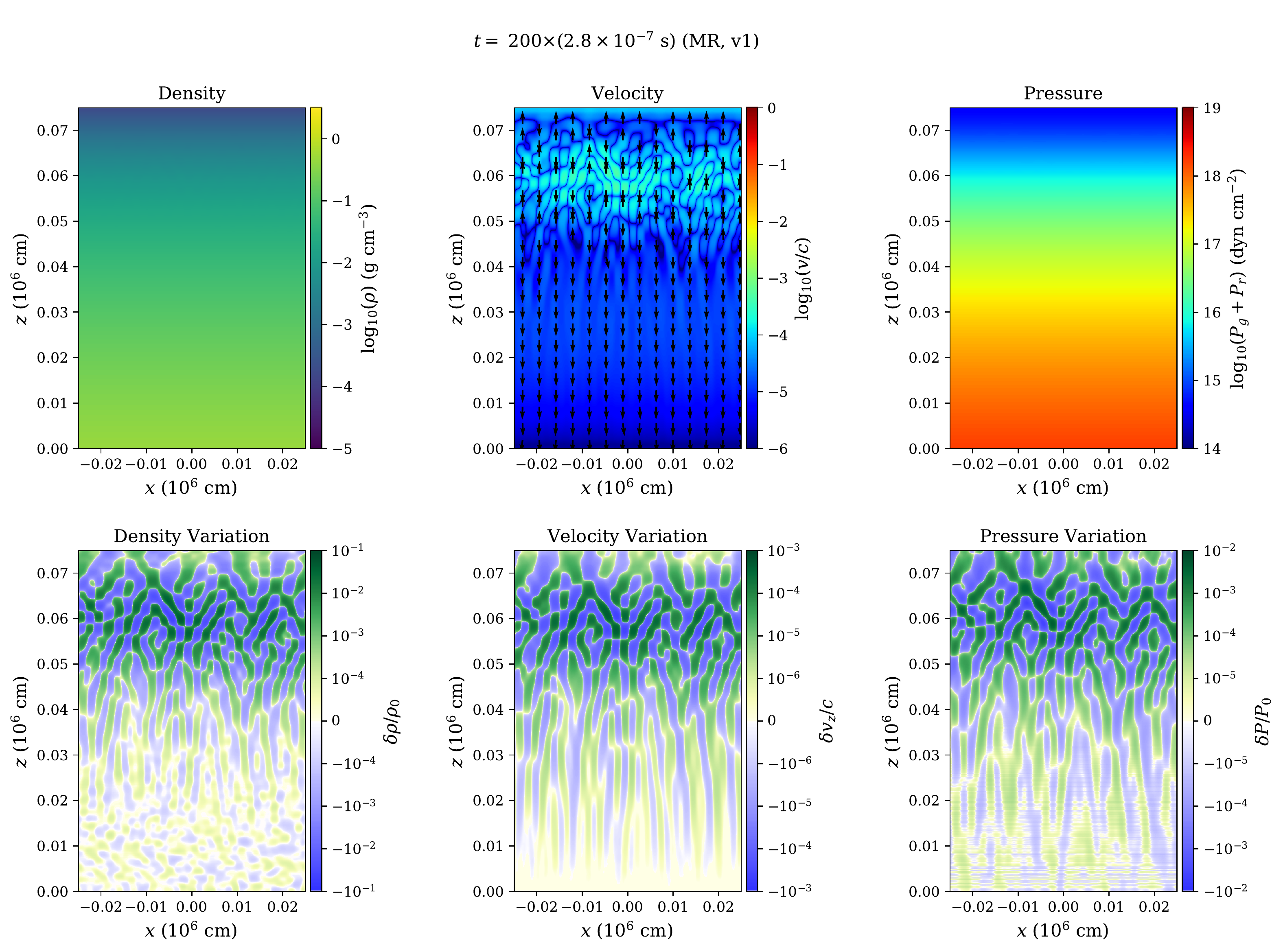}
    \caption{Snapshot at $t=200\times(2.8\times10^{-7}\ \mathrm{s})$ in the \lz{MR} simulation.}
    \label{fig:snapshot_200}
\end{figure*}

\begin{figure*}
	\includegraphics[width=0.82\textwidth]{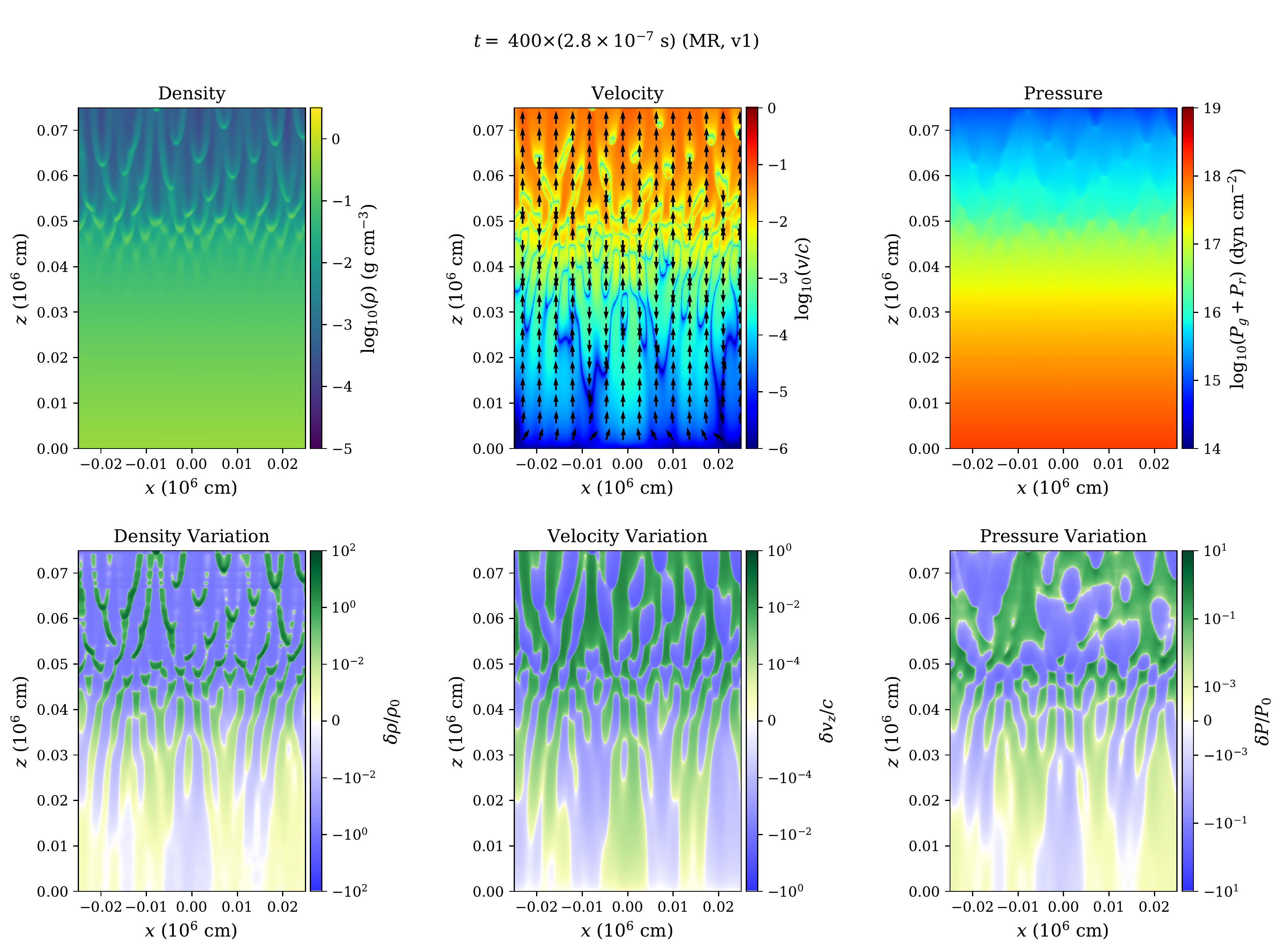}
    \caption{Snapshot at $t=400\times(2.8\times10^{-7}\ \mathrm{s})$ in the \lz{MR} simulation.}
    \label{fig:snapshot_400}
\end{figure*}

\begin{figure*}
	\includegraphics[width=0.82\textwidth]{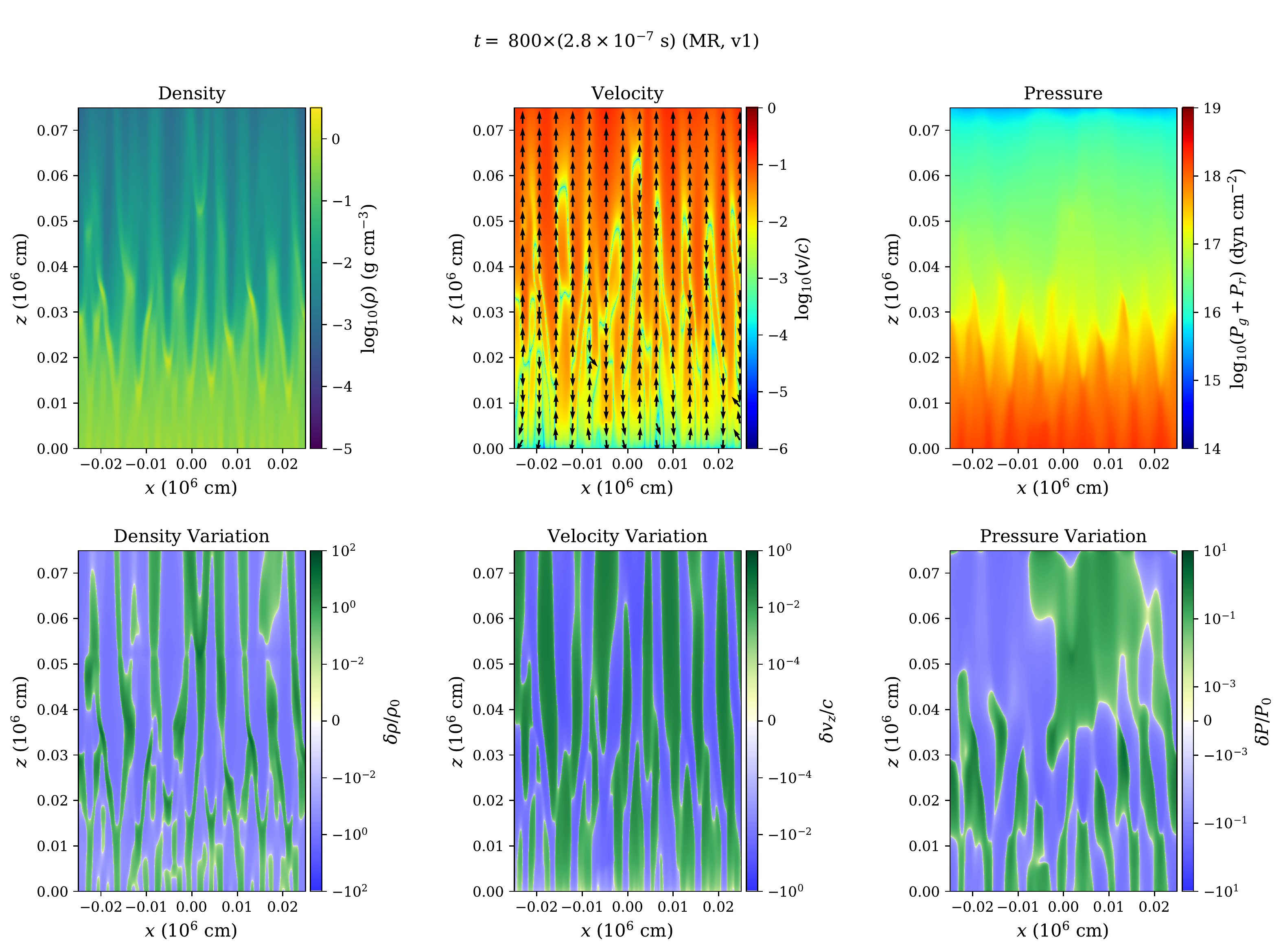}
    \caption{Snapshot at $t=800\times(2.8\times10^{-7}\ \mathrm{s})$ in the \lz{MR} simulation.}
    \label{fig:snapshot_800}
\end{figure*}

\begin{figure*}
	\includegraphics[width=0.8\textwidth]{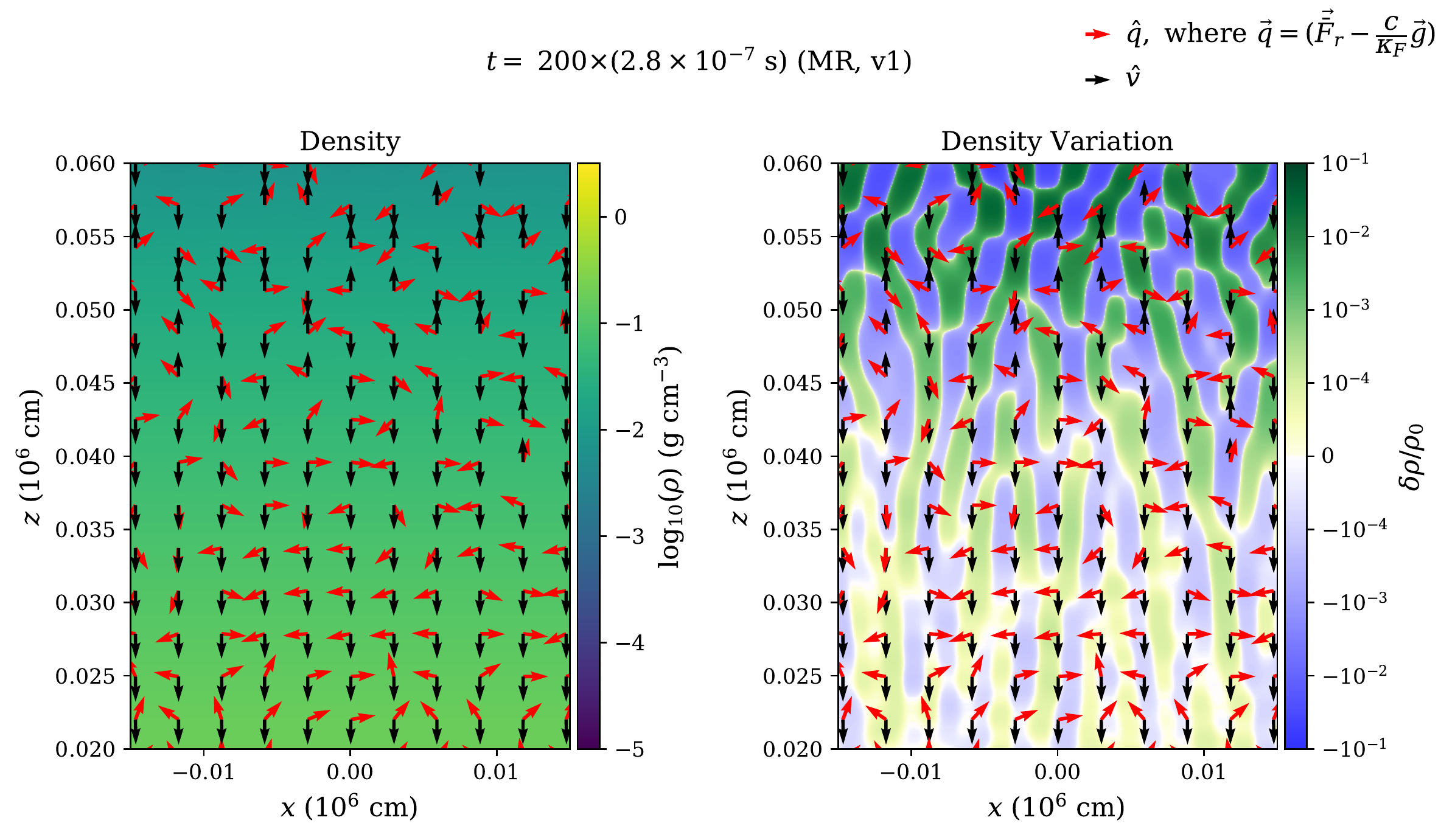}
    \caption{Zoom-in profiles at $t=200\times(2.8\times10^{-7}\ \mathrm{s})$ in the \lz{MR} simulation.  Black arrows indicate unit vectors in the direction of the local velocity, and red arrows indicate unit vectors in the direction of the portion of the radiation flux that is not providing hydrostatic support.}
    \label{fig:snapshot_200_zoom_in}
\end{figure*}

\begin{figure*}
	\includegraphics[width=0.8\textwidth]{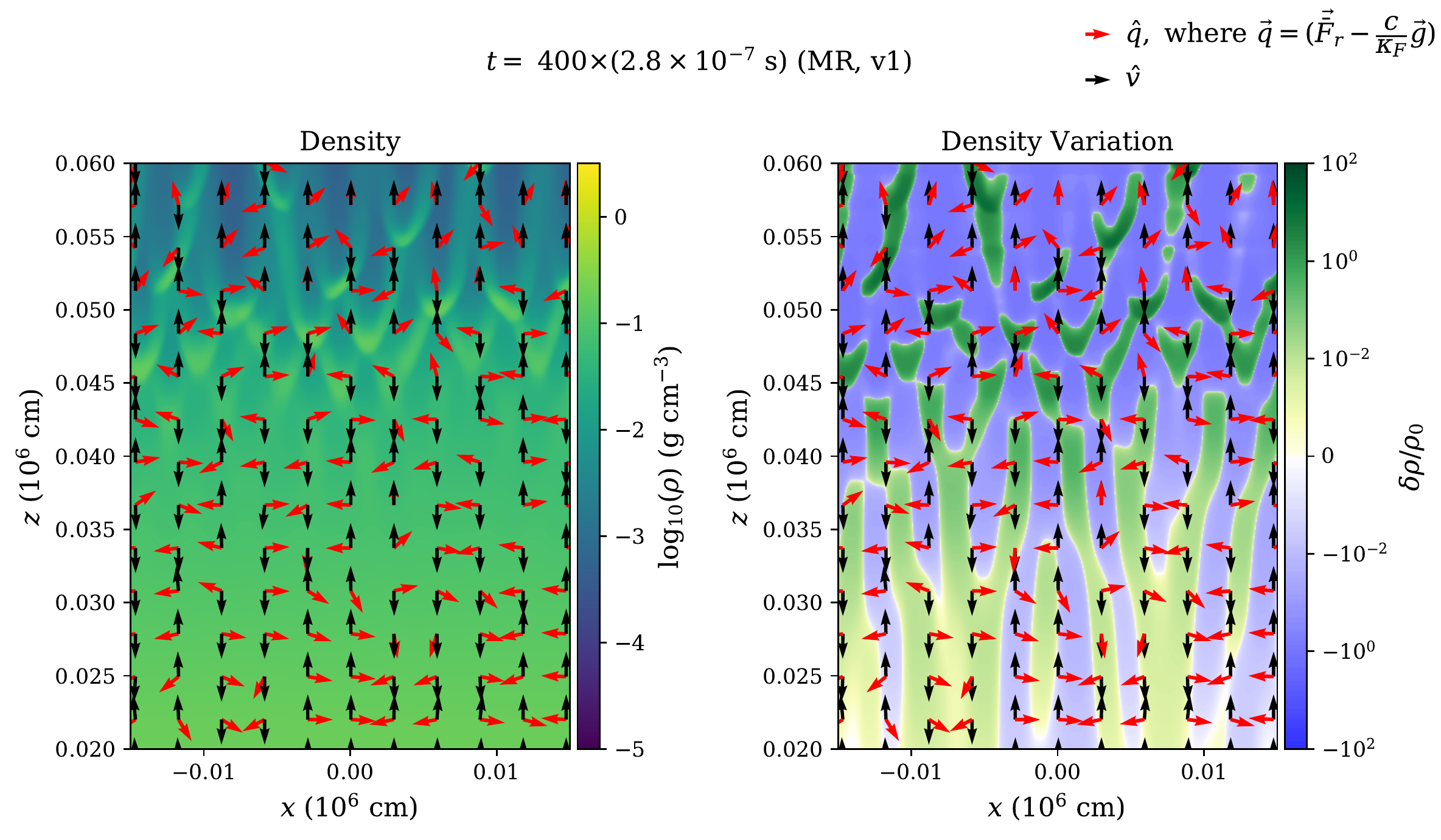}
    \caption{Zoom-in profiles at $t=400\times(2.8\times10^{-7}\ \mathrm{s})$ in the \lz{MR} simulation.}
    \label{fig:snapshot_400_zoom_in}
\end{figure*}

\begin{figure*}
	\includegraphics[width=0.8\textwidth]{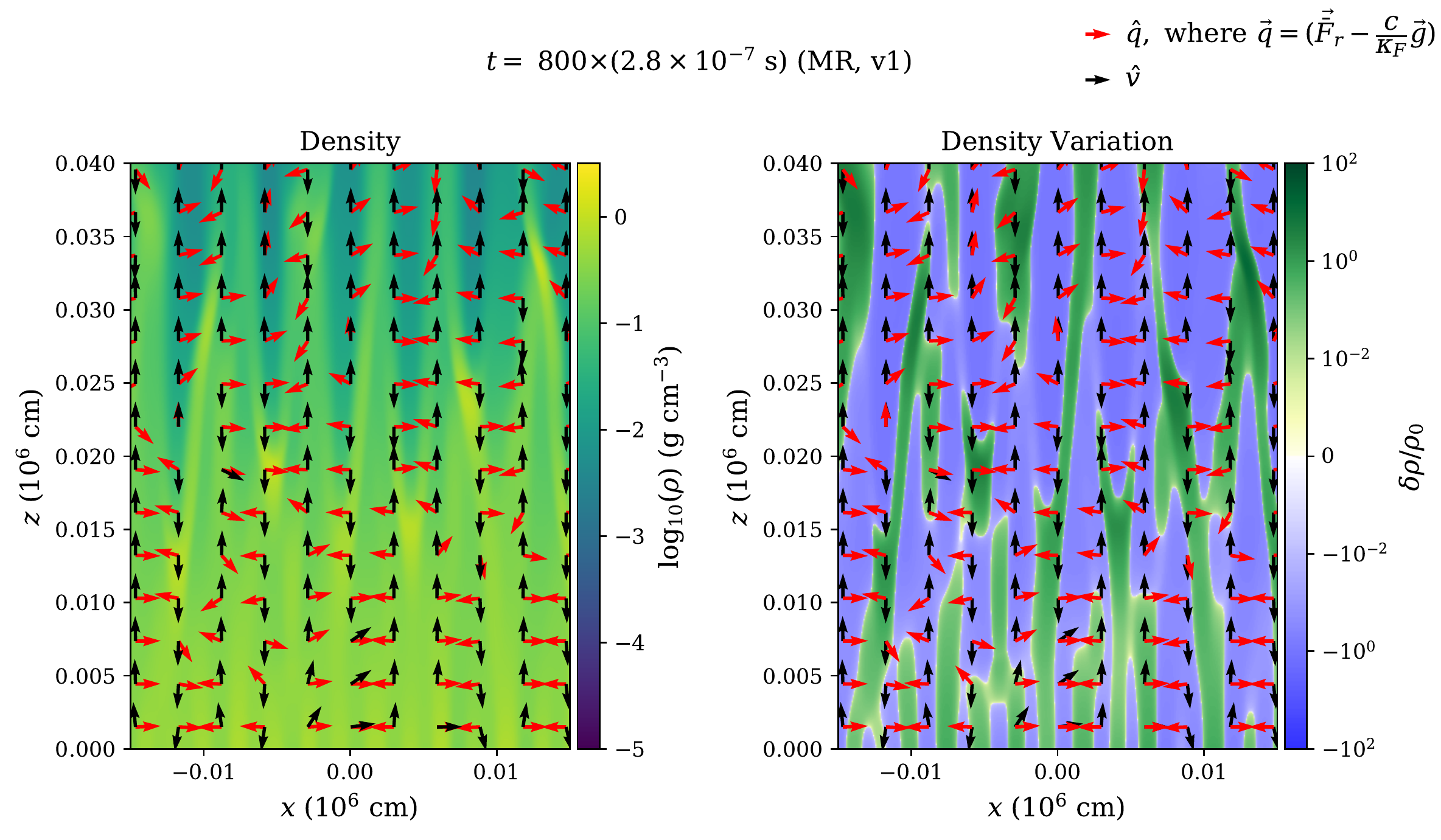}
    \caption{Zoom-in profiles at $t=800\times(2.8\times10^{-7}\ \mathrm{s})$ in the \lz{MR} simulation.}
    \label{fig:snapshot_800_zoom_in}
\end{figure*}

We pick three snapshots at $t/t_{\mathrm{sim}}=200$, $400$ and $800$ from the \lz{MR} simulation. The format of each snapshot is the same, where the upper-left panel refers to the gas density; lower-left refers to the gas density variation; upper-middle refers to the fluid velocity; lower-middle refers to the vertical velocity variation; upper-right refers to the gas-radiation pressure; lower-right refers to the gas-radiation pressure variation. Note that in the upper-middle panel, the color represents the magnitude of the velocity and the black arrows indicates the directions of the velocity.

As shown in \autoref{fig:snapshot_200}, the system is at the early stage of the linear instability at $t/t_{\mathrm{sim}}=200$ before the first synchronization time and the instability grows independently at each height. Although the density profile is rather smooth, the instability patterns are already clear and roughly consistent in the variation (lower) panels.  Moreover, we can find that the tilted angles of the instability patterns become more vertical towards the bottom. This agrees with our analytical solution that the maximum growth rate peaks at more vertically tilted angles as the diffusion regime becomes slower (i.e. $M_0 \rightarrow 0$) and we will discuss details in the \hyperref[sec:sanity_check]{Section 4.2}. As shown in \autoref{fig:snapshot_400}, the contrast in density profile at the top is large enough to see the pattern. Meanwhile, the instability patterns become more irregular at $t/t_{\mathrm{sim}}=400$. The instabilities at higher regions clearly step into the nonlinear phase, which would gradually propagate down and interfere the regions below. Furthermore, the gas is sinking after losing the radiation support because of the spatial decoupling between gas and radiation. Eventually, the whole atmosphere collapses at $t/t_{\mathrm{sim}}=800$ as shown in \autoref{fig:snapshot_800}. At this stage, the gas and radiation are driven into different channels by the instabilities, where the gas sinks downwards and the radiation escapes upwards.

Now we can zoom in to these selected snapshots to track the gas-radiation decoupling via gas velocity and heat flow. \autoref{fig:snapshot_200_zoom_in} shows the density distribution (left) and density variation (right) at time $t/t_{\mathrm{sim}}=200$, when we are still in the linear regime.  Arrows show the directions of velocity (black) and the portion of the radiation flux (red) that is not providing hydrostatic support.  As we discussed in \hyperref[sec:pbi_overview]{Section 2}, the vertical component of this portion is what drives the linear photon bubble instability, and indeed we find that this heat predominantly flows from regions of high perturbed density to the regions of low perturbed density (see red arrows in \autoref{fig:snapshot_200_zoom_in}).  This is a clear indication of photon bubble instability in the slow diffusion regime. Since the heat flow is mostly in the form of radiation, the radiation further evacuates the low perturbed density regions and eventually causes spatial decoupling from the gas. Meanwhile, the gas in high perturbed density regions is gradually sinking because of loss of radiation pressure support (see black arrows in \autoref{fig:snapshot_200_zoom_in}). This decoupling process becomes more intense as the instability grows. As shown in \autoref{fig:snapshot_400_zoom_in} at $t/t_{\mathrm{sim}}=400$, the instability in upper region grows into nonlinear phase and the density contract become more clear. The gas in high density regions keeps sinking, while the radiation in low density regions escapes upwards and blows the gas away. Finally, the whole system become nonlinear at $t/t_{\mathrm{sim}}=800$ as shown in \autoref{fig:snapshot_800_zoom_in}, where the gas and radiation are clearly separated in different channels by the effects of photon bubble physics.  Note that by this time the fluid velocity is no longer entirely vertical at the base, indicating significant bending of the magnetic field lines there.

\subsection{Comparison with Linear Theory}
\label{sec:sanity_check}

\begin{table*}
	\centering
	\begin{tabular}{lcccccccccccccc}
		\hline
		$z$ & $M_0$ & $h$ & $h_g$ & $l_{\mathrm{vis}}$ & \multicolumn{2}{c}{LR (v0)} & \multicolumn{2}{c}{MR (v1)} & \multicolumn{2}{c}{HR (v2)} & \multicolumn{2}{c}{SHR (v3)} & \multicolumn{2}{c}{UHR (v4)}\\
	    & & & & & \multicolumn{2}{c}{$t/t_{\mathrm{sim}}=300$} & \multicolumn{2}{c}{$t/t_{\mathrm{sim}}=250$} & \multicolumn{2}{c}{$t/t_{\mathrm{sim}}=150$} & \multicolumn{2}{c}{$t/t_{\mathrm{sim}}=100$} & \multicolumn{2}{c}{$t/t_{\mathrm{sim}}=100$}\\
		& & & & & $\theta_{\mathrm{max}}$ & $\lambda$ & $\theta_{\mathrm{max}}$ & $\lambda$ & $\theta_{\mathrm{max}}$ & $\lambda$ & $\theta_{\mathrm{max}}$ & $\lambda$ & $\theta_{\mathrm{max}}$ & $\lambda$\\
		($R_{\star}$) & & ($\mathrm{cm}$) & ($\mathrm{cm}$) & ($\mathrm{cm}$) & ($\mathrm{deg}$) & ($\mathrm{cm}$) & ($\mathrm{deg}$) & ($\mathrm{cm}$) & ($\mathrm{deg}$) & ($\mathrm{cm}$) & ($\mathrm{deg}$) & ($\mathrm{cm}$) & ($\mathrm{deg}$) & ($\mathrm{cm}$)\\
		\hline
		0.02 & 0.01 & 19760 & 148 & 433 & 87.36 & 7116 & 87.13 & 4277 & 86.44 & 1661 & 84.94 & 622 & 85.17 & 644 \\
		0.03 & 0.03 & 16438 & 123 & 506 & 85.98 & 8102 & 85.52 & 4202 & 84.13 & 1505 & 82.17 & 660 & 82.17 & 660 \\
		0.04 & 0.08 & 13133 & 98 & 568 & 82.86 & 6185 & 82.05 & 3804 & 80.31 & 1739 & 77.85 & 873 & 77.15 & 777 \\
		0.05 & 0.29 & 9887 & 73 & 843 & 76.32 & 6107 & 74.78 & 3636 & 71.90 & 1673 & 67.74 & 712 & 68.61 & 861 \\
		\hline
	\end{tabular}
	\caption{\lz{Data measurements in linear phase of all simulations}}
	\label{tab:sim_measure}
\end{table*}

\begin{figure}
	\includegraphics[width=\columnwidth]{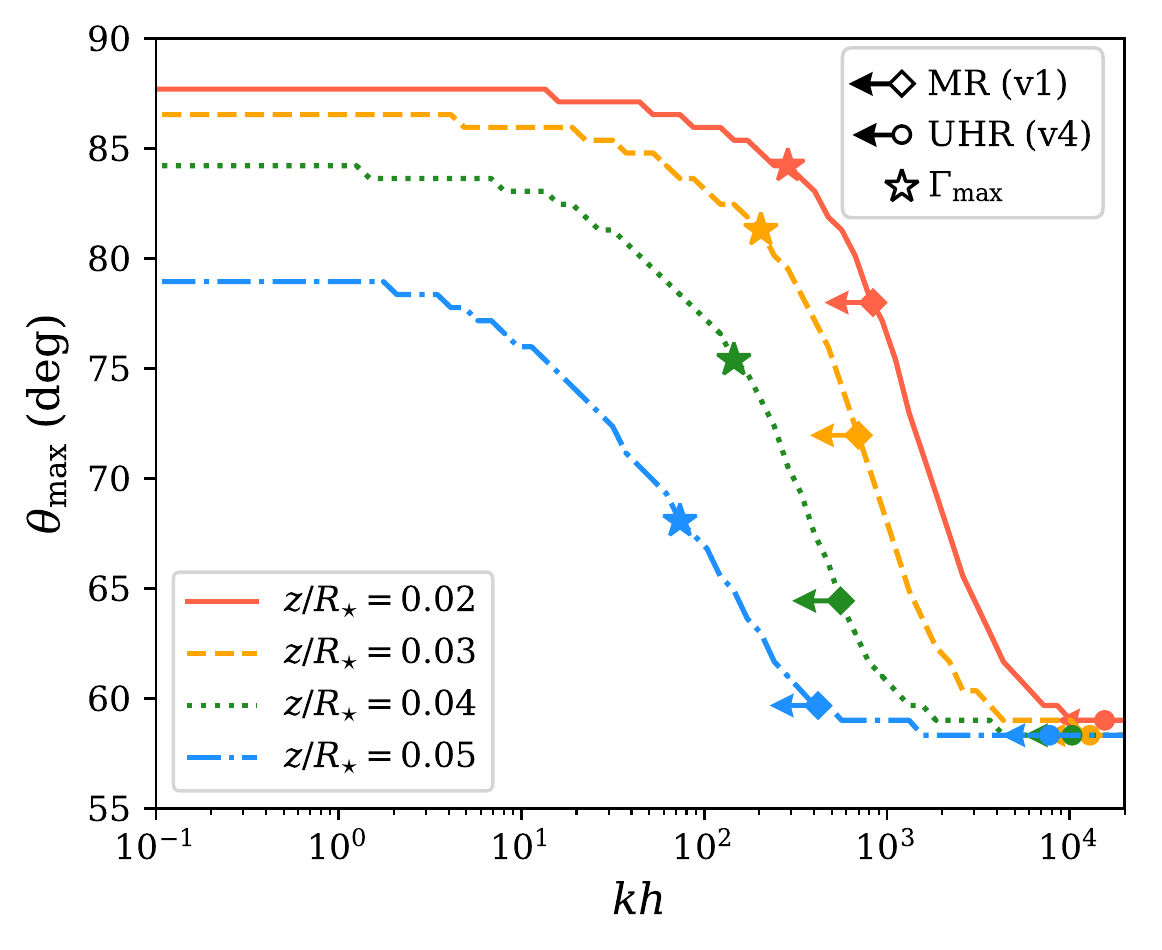}
    \caption{The analytical dependence of the angle corresponding to the maximum growth rate ($\theta_{\max}$) as a function of wavenumber. \lz{The stars indicate the theoretical maximum growth rate at each height. The diamonds and dots with arrows refer to the minimum resolved wavelength (2 grid zones) in the MR and UHR simulations, respectively.} }
    \label{fig:angle_max}
\end{figure}

At each wavelength, there is a specific angle $\theta_{\mathrm{max}}$ between the wave vector direction and the vertical that maximizes the linear growth rate of the instability (see \autoref{fig:growthRate3D}). This is essential for understanding the instability pattern. Therefore, we solve for $\theta_{\mathrm{max}}$ at each selected height as a function of wavelength, and show the results in \autoref{fig:angle_max}.  \lz{The stars indicate the maximum growth rate by numerically solving the dispersion relation (\autoref{eq:dispersion_blaes_norm}) given the profile at each height.} The shortest wavelengths that can be resolved (2 grid zones for a crest and a trough) are marked with \lz{diamonds for the \lz{MR} simulation and dots for the \lz{UHR} simulation, where the arrow indicates the longer wavelengths that can be resolved.}  Short wavelengths generally grow fastest \lz{until the viscous length scale is reached} (\autoref{fig:growthRate3D}), and therefore tend to dominate the instability pattern. High altitudes have the smallest $\theta_{\mathrm{max}}$, and $\theta_{\mathrm{max}}$ approaches $90^{\circ}$ as the height decreases. This is consistent with the overall instability patterns becoming less tilted with decreasing height as shown in \autoref{fig:snapshot_200}. In \autoref{fig:angle_max}, we also notice the wider range of $\theta_{\mathrm{max}}$ in higher altitude regions. This suggests the `block structure' along these tilted patterns from the mixing modes because some less dominant modes tweak the substructure of the tilted pattern to be more vertical, which can also be seen directly in the variation (lower) panels of \autoref{fig:snapshot_200}. 

\begin{figure}
	\includegraphics[width=\columnwidth]{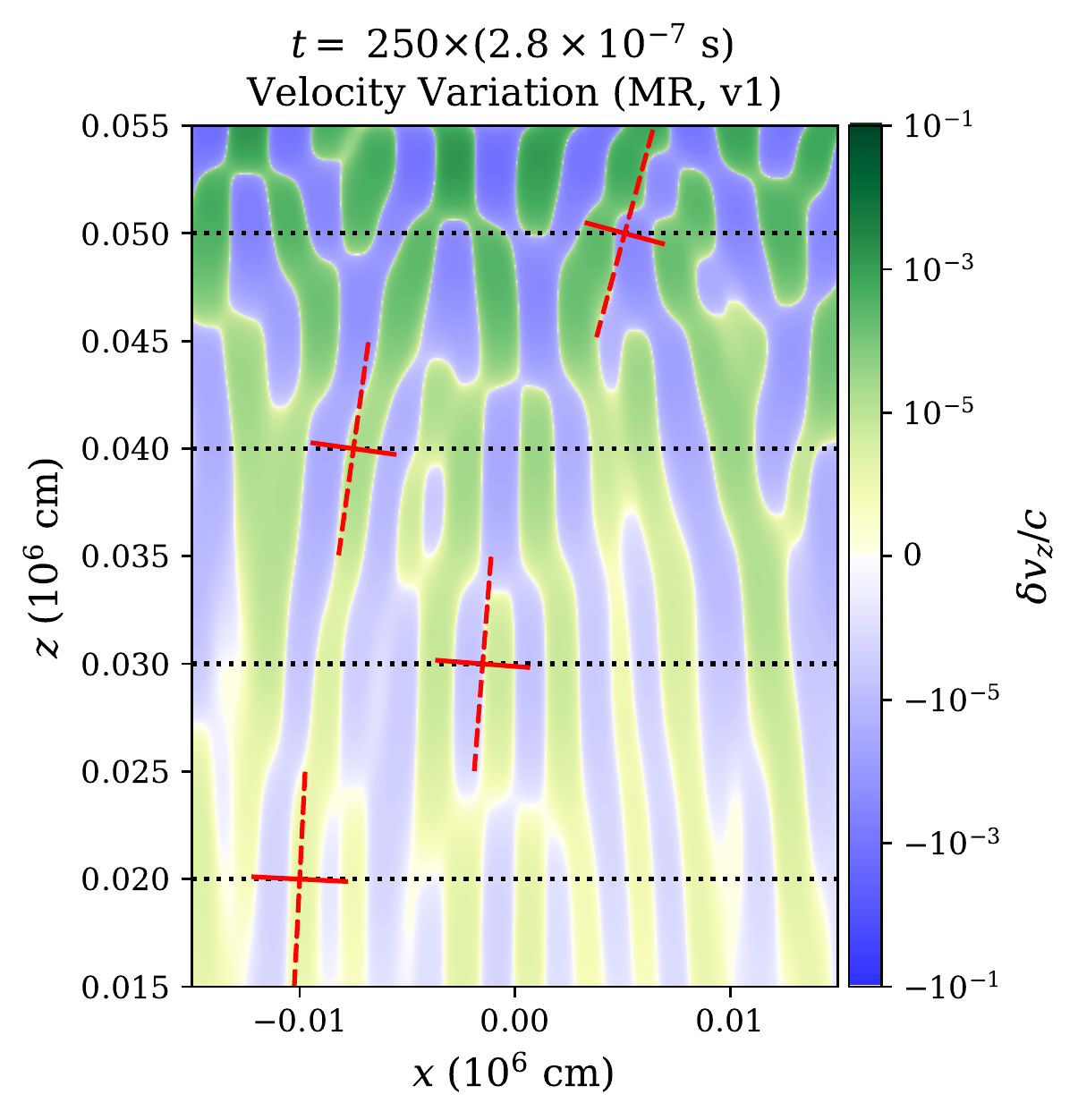}
    \caption{
    Comparison of the tilt angles of the dominant modes at different heights in the \lz{MR} simulation with that expected from linear theory.  Dashed lines are oriented along constant phase surfaces, and solid lines indicate wavelength perpendicular to the constant phase surfaces.}
    \label{fig:snapshot_200_pattern}
\end{figure}

To test the analytical expectation of the instability growth rate and the tilted angles, we first need to measure the wavelength from the fluctuations in the linear growth phase. Since the modes are mixed, we distinguish the wavelengths in Fourier space. Here, we briefly describe how we obtain the dominant wavelength from the simulation data. At each selected height, we first analyze the horizontal profile to obtain the perturbation with respect to the unperturbed initial condition (e.g. $\delta \rho$, $\delta v_z$ or $\delta P$) as a function of horizontal distance $x$.  Next, we project these variables into Fourier space, in order to identify the dominant mode from the peak in the power spectrum. We then pick the measured wavelength of the dominant mode and calculate the corresponding $\theta_{\mathrm{max}}$. Note that here the horizontal wavelength $\lambda_{\mathrm{\perp}} = \lambda/\sin{\theta_{\mathrm{max}}}$ (where `$\perp$' means perpendicular to the magnetic field) is the horizontal projection of the real wavelength ($\lambda$). Therefore, we need to correct it with the angle by iterating the value of $\theta_{\mathrm{max}}$ until it converges within some tolerance we choose (e.g. $10^{-8}$). The results of the measurements based on $\delta v_z$ in \lz{all} simulations are listed in \autoref{tab:sim_measure}. As shown in \autoref{fig:snapshot_200_pattern}, we plot the measured wavelengths (solid red lines) and the predicted constant phase plane orientations (dashed red lines) of the dominant mode at each selected height, which are quite consistent with the instability pattern observed in the simulation. However, we notice that the fastest growing wavelength at each height is not what we expected earlier as 2 grid zones in \lz{the MR simulation} but it is $\sim 50$ grid zones instead. This is because of damping effects arising from numerical diffusion. In all versions of our photon bubble simulations, these numerical damping effects start to become important at wavelengths below $\sim 50$ grid zones.  This can be contrasted with the {\it ZEUS} simulations of \citet{2005ApJ...624..267T} of photon bubbles in the rapid diffusion regime, where numerical damping was significant at wavelengths below $\sim 10$ grid zones.
Thus, the numerical damping at short length scales makes resolution a challenge.

\begin{figure}
	\includegraphics[width=\columnwidth]{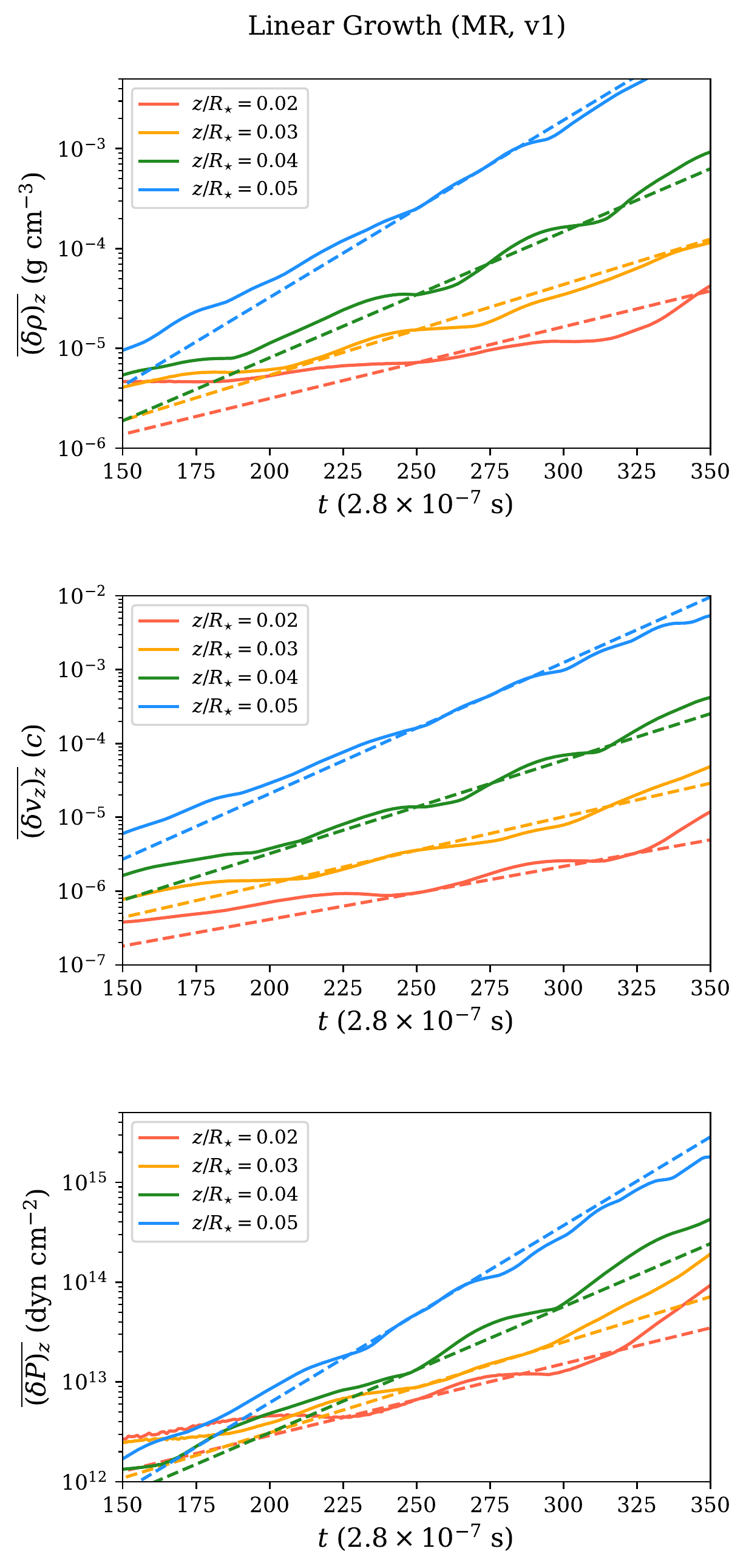}
    \caption{
    Comparison of linear growth rates with behavior measured in the \lz{MR} simulation.}
    \label{fig:growth_rate_check}
\end{figure}

\begin{figure}
	\includegraphics[width=\columnwidth]{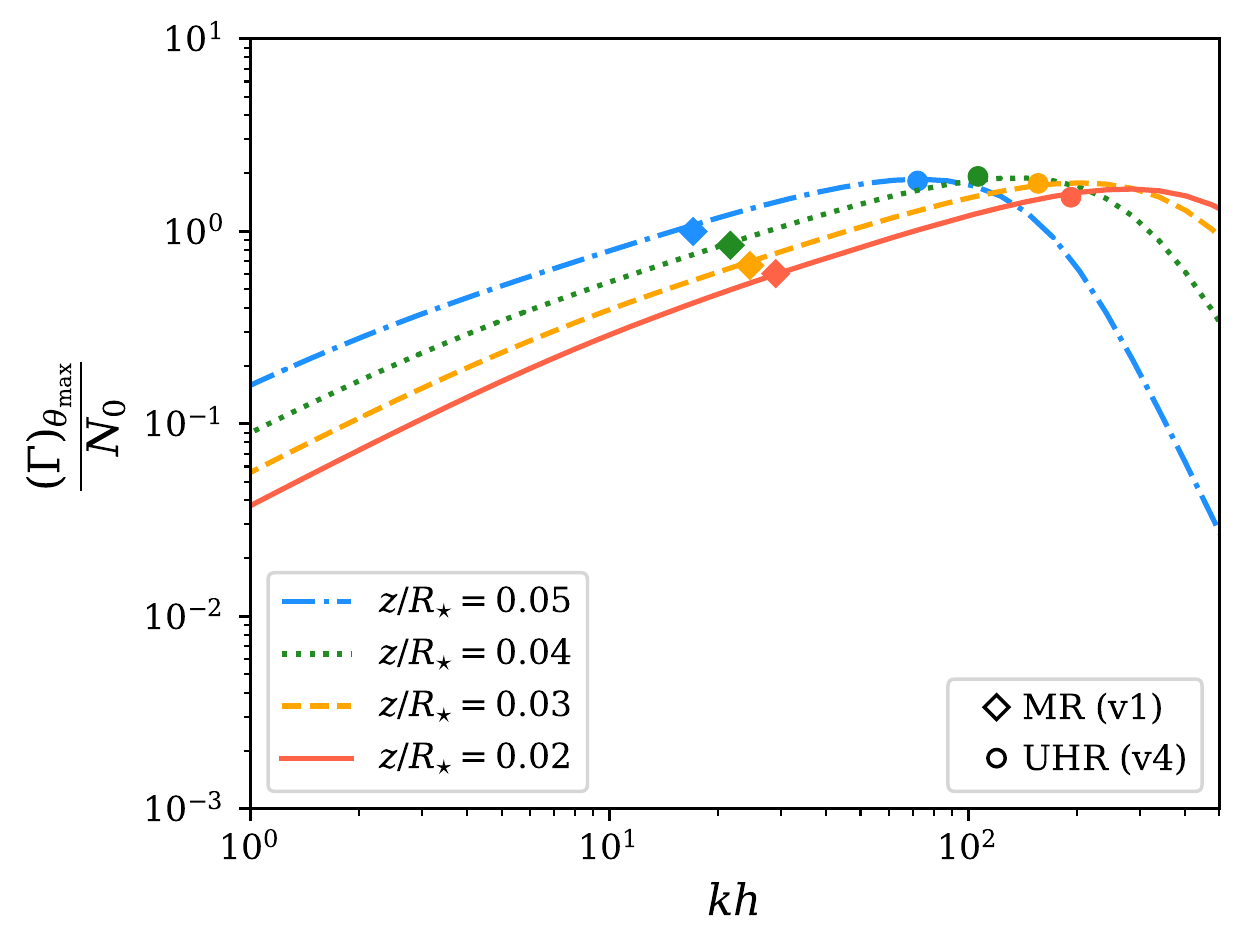}
    \caption{The analytical dependence of the growth rate, \lz{scaled with the sound crossing frequency over a scale height $N_0=c_r/h$,} at $\theta_{\mathrm{max}}$ on wavenumber. \lz{The diamonds and dots represent the linear fits to the \lz{MR} and \lz{UHR} simulation data, respectively, given the measured wavelengths of the dominant modes.}}
    \label{fig:growth_rate_fit}
\end{figure}

With the measurements of the dominant wavelength at each height, we can compute the analytical growth rates and compare them with the simulation data as illustrated in \autoref{fig:growth_rate_check}.  \lz{In the MR simulation,} the measured linear growth rates (solid lines) are fairly consistent with the analytical calculation (dashed lines), where the analytical calculation is based on the snapshot at $t/t_{\mathrm{sim}}=250$. Similarly, we also compute the analytical growth rate in the \lz{MR} and \lz{UHR} simulations as a function of wavelength by \lz{selecting the angle $\theta_{\mathrm{max}}$ corresponding to the maximum growth rate. }
As shown in \autoref{fig:growth_rate_fit}, each lines represent the analytical solutions at different heights. \lz{The solid diamonds and dots are the measurements of the dominant modes in the \lz{MR} and \lz{UHR} simulations respectively.} The linear growth rates are generally larger in higher altitude regions simply because the radiation diffuses faster. The simulated growth rates of the dominant modes again are fairly consistent with the analytical solution. \lz{Note that the growth rate reaches a finite maximum and then declines toward shorter wavelengths because of radiation viscosity.  The length scale of this peak growth rate is small and requires high resolution in the simulation. Among the five simulations, only the SHR and UHR simulations roughly reach the analytical maximum growth rate of the photon bubble instability. Details will be discussed in the next section. }

\subsection{Resolution Dependence}
\label{sec:resolution_dep}

\begin{figure*}
	\includegraphics[width=1.0\textwidth]{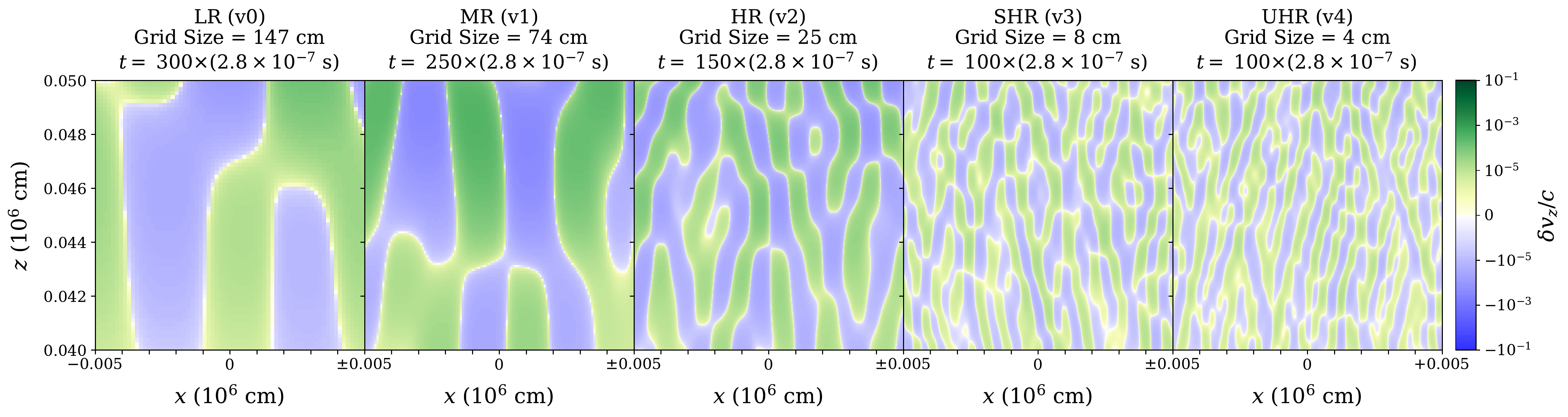}
    \caption{
    \lz{Side-by-side comparison of linear-phase photon bubble instability at different resolutions, for the same portions of the simulation domains.  The length scales of the dominant modes decrease with increasing resolution, until convergence is achieved at the viscous length scale for the two highest resolution simulations on the right.}
    }
    \label{fig:res_compare}
\end{figure*}

\begin{figure}
	\includegraphics[width=1.0\columnwidth]{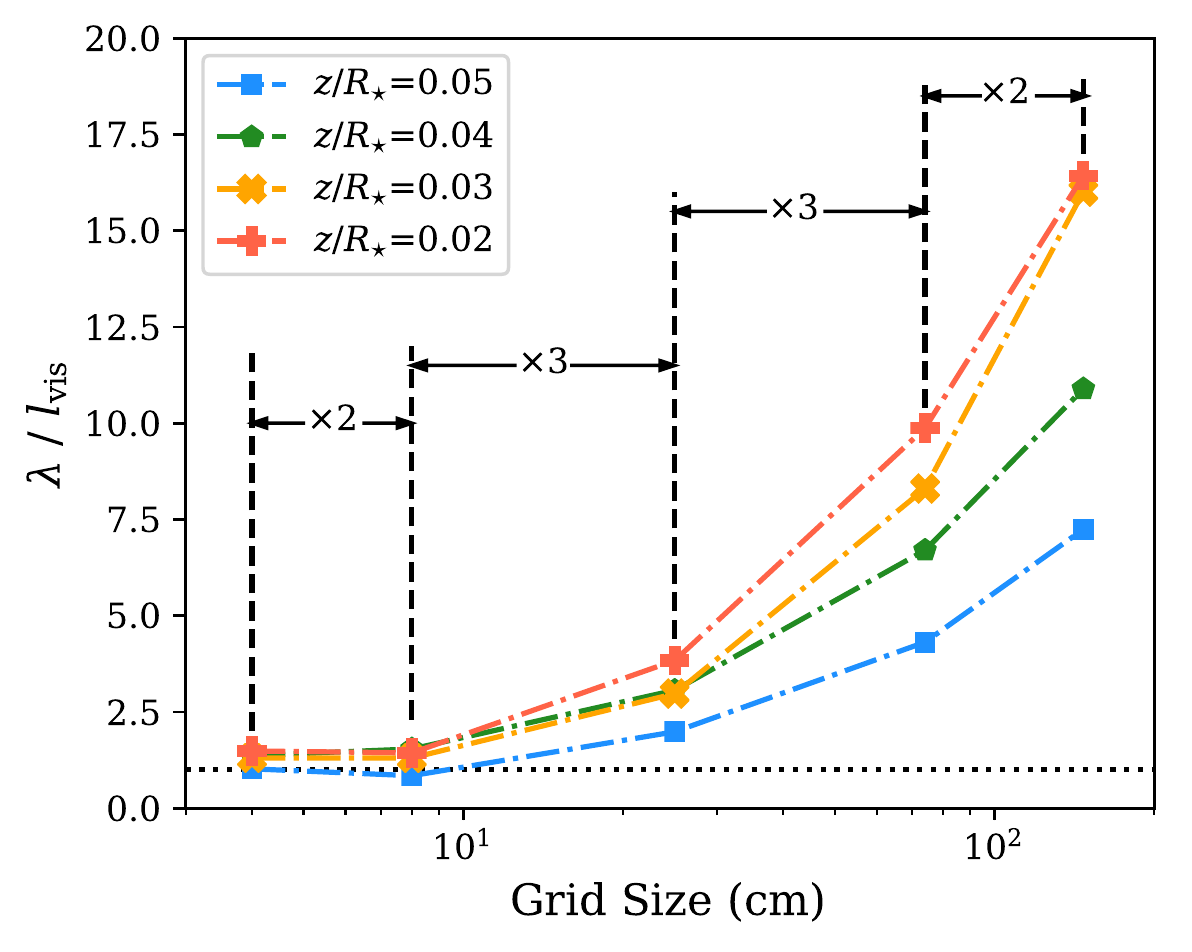}
    \caption{
    \lz{Numerical convergence of the different resolution simulations, at different altitudes as indicated.  The horizontal intervals $\times2$ and $\times3$ indicate the increased resolution factors between simulations.  The horizontal dotted line indicates unity, and shows that the two highest resolution simulations have dominant wavelengths equal to the viscous length scale for maximum growth, in agreeement with linear theory.}
    }
    \label{fig:res_converge}
\end{figure}

\begin{figure*}
	\includegraphics[width=1.0\textwidth]{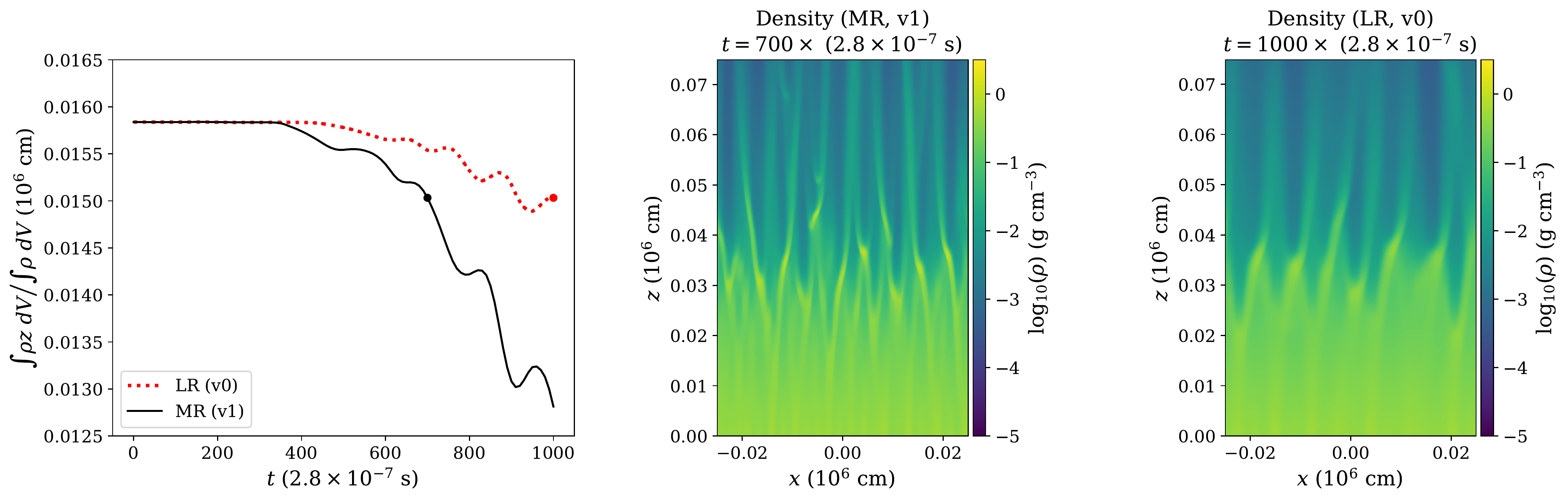}
    \caption{
    Left:  Evolution of mass-weighted height of the atmosphere in the high resolution (black) and low resolution (red dotted) simulations.  Snapshots of the density distribution a times indicated by the points in the left panel (when both simulations have collapsed to the same height) are shown in the middle (high resolution) and right (low resolution) panels.}
    \label{fig:timestream2}
\end{figure*}

\begin{figure*}
	\includegraphics[width=1.0\textwidth]{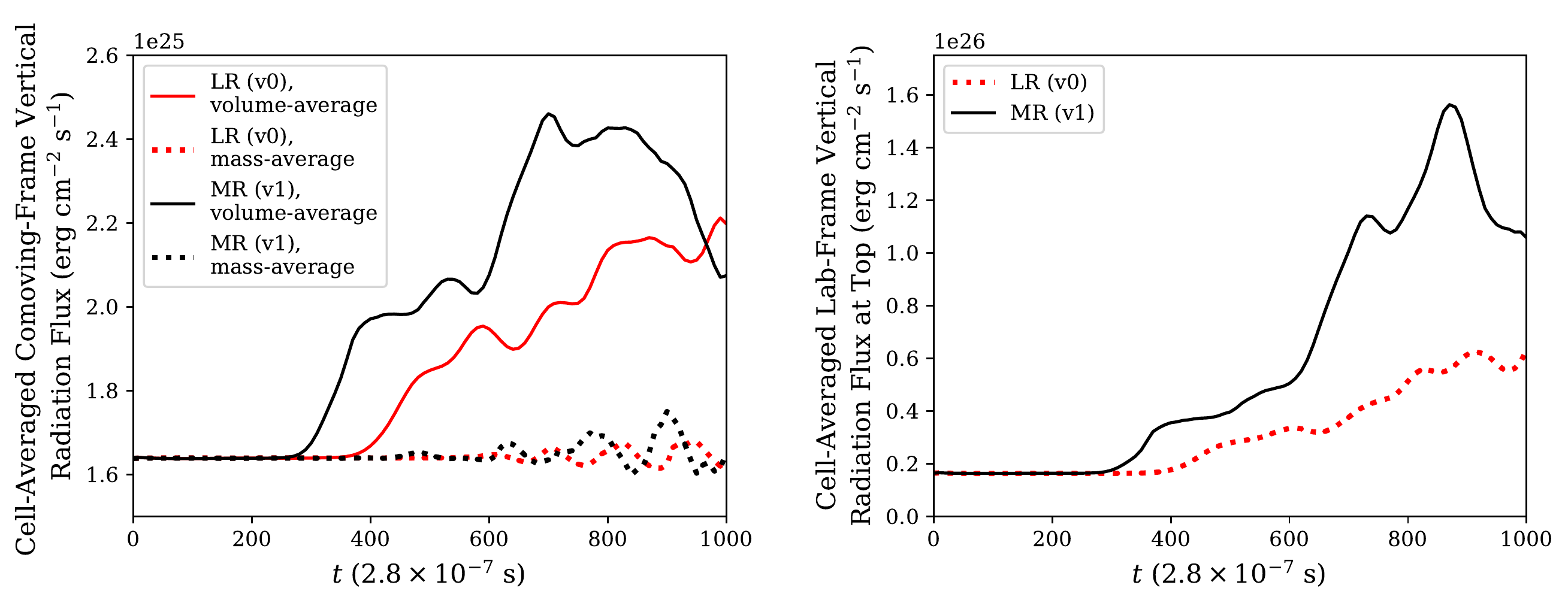}
    \caption{
    Left: Evolution of volume-averaged (solid) and mass-averaged (dotted) comoving-frame vertical radiation fluxes in the high resolution (black) and low resolution (red) simulations.  Right:  Evolution of lab-frame vertical radiation flux leaving through the top of the simulation domain.}
    \label{fig:timestream1}
\end{figure*}

According to the linear theory (see \hyperref[sec:derivation_pbi]{Appendix A} for details), we expect that simulations of photon bubble instability will be resolution-dependent because the instability grows faster at shorter wavelengths \lz{until the wavelength reaches the maximum at radiation viscous length scale, which is small and thus requires high resolution}. The shortest wavelength that can be resolved is at least two grid zones in order to resolve a crest and a trough, but numerical diffusion affects the results at much longer wavelengths than this. As we just discussed, we find that unstable modes roughly require at least $50$ grid zones to avoid significant numerical damping in our \lz{MR} resolution simulation. In the \lz{LR} simulation, we perform similar measurements at $t/t_{\mathrm{sim}}=300$ as shown in \autoref{tab:sim_measure}, where the fastest growing wavelengths are on average $47$ grid zones, comparable to what we found in the \lz{MR} simulation. Note that we select a later snapshot compared with the \lz{MR} simulation because the dominant modes in the \lz{LR} simulation have longer wavelength in general, and therefore grow more slowly.  \lz{This trend continues as we keep increasing the resolution, until the simulation starts to resolve the radiation viscous length scale. } The discrepancy of the growth rates at different heights increases when the dominant modes move toward longer wavelengths as shown in \autoref{fig:growth_rate_fit}, which indicates that the instability in the low altitude regions is suppressed and synchronized faster by the instability propagating downward from high altitudes. This complicates the mode analysis at the low altitudes. In short, simulations of photon bubbles need to have high enough resolution that the more slowly growing modes at low altitude are at least able to start their linear growth phase before they are affected by downward propagation of the faster growing modes at high altitude.  On the other hand, we find that the nonlinear outcome of both the \lz{LR} and \lz{MR} simulations are qualitatively similar, and they both collapse.  However, the \lz{LR} simulation takes longer to collapse, and the photon bubble channels that form have longer horizontal length scales. 

\lz{In order to study the resolution dependence of photon bubble instability, we ran three extra simulations through the linear growth phase by increasing the grid cell size to 25~cm, 8~cm and 4~cm, respectively. Snapshots of the instability patterns of all five simulations are shown in \autoref{fig:res_compare}. The dominant wavelength keeps decreasing as the resolution increases until the SHR simulation at resolution $\sim 8$~cm.  A further increase in resolution to $\sim4$~cm in the UHR simulation produces hardly any change in the spatial scale of the photon bubble. As shown in \autoref{fig:res_converge}, we have roughly reached the convergence at grid cell size $\sim8$ cm, where the dominant wavelength nearly reaches the radiation viscous length scale at each selected height. }

The evolution in the nonlinear regime is also affected by resolution, as we show in \autoref{fig:timestream2}. The left panel shows that the mass-weighted height of the atmosphere decreases as the system collapses.  The \lz{MR} simulation collapses more rapidly because it is able to resolve the faster growing shorter wavelength modes of the instability.  These modes persist well into the nonlinear regime, as illustrated in the density snapshots shown in the middle and right hand panels of \autoref{fig:timestream2}. By simply counting the peaks in the horizontal density variation at each height in the snapshot, we find that the average horizontal length scale in the snapshots are 3375~cm for the \lz{MR} simulation and 6168~cm for the \lz{LR} resolution simulation, \lz{which is consistent with the horizontal projection of the photon bubble wavelength in the high altitude regions}.  The ratio of 1.8 is close to the factor of two difference in grid resolution.

\autoref{fig:timestream1} depicts the evolution of the vertical radiation flux in the two simulations.  The left hand panel shows the vertical radiation flux averaged over the simulation domain, with solid curves showing a volume average and dotted curves showing a mass-weighted average.  The latter is far below the former, illustrating the fact that photons are escaping preferentially through the low density channels formed by the photon bubble instability, and therefore the denser regions are no longer supported against gravity by radiation pressure, causing the collapse.  The radiation escape rate is significantly higher for the \lz{MR} simulation, which is why it collapses faster. The right panel shows the horizontally-averaged radiative flux that leaves through the top of the simulation domain. Again, the rate at which radiation leaves the simulation domain is significantly higher for the \lz{MR} simulation. \lz{Note that we did not evolve the HR, SHR and UHR simulations to the nonlinear regime because of limited computational resources. }

\lz{The closest simulations to those we have presented here are those of \citet{1997ApJ...478..663H}, who also simulated the growth of photon bubbles in static atmospheres. Because of better computational resources available today, we have been able here to run simulations at much higher resolutions and use a more extended simulation domain. Even our lowest resolved simulation (LR) has grid cell sizes smaller by factor of $\sim2.4$ than the finest grid used in \citet{1997ApJ...478..663H}. They also used grids that reduced resolution with altitude, which would cause slower growth of the photon bubble instability in these regions. Our simulations cover a broader parameter space of radiation diffusion, with $M_0$ ranging from $\sim0.005$ to $\sim100$. Our simulations also directly solve the full angle-dependent radiative transfer equation, whereas they used flux-limited diffusion. Therefore, we are able to reach length scales where photon viscosity is important, and even use grid cell sizes that are optically thin.  The maximum growth rate of our photon bubbles is naturally limited by the radiation viscosity at high enough resolution. The magnetic field direction in our simulations is vertical (the same as the gravitational field) and thus our simulations have a left-right horizontal symmetry in the fastest growing photon bubble modes.} \lz{\citet{1997ApJ...478..663H} adopted a slightly tilted magnetic field instead, which breaks this symmetry of their photon bubbles, and the dominant modes only propagate in one direction. Finally, we solve the MHD equations so the gas motion is constrained by a dynamical magnetic field ($B=10^{10}$~Gauss), while \citet{1997ApJ...478..663H} impose 1D motion of the gas to mock up the effects of a strong magnetic field.}

\lz{Despite these differences, the numerical outcomes in both sets of simulations are very similar. Both we and \citet{1997ApJ...478..663H} find that photon bubbles align close to the equilibrium magnetic field in the lower altitudes where diffusion is slowest, and that in the nonlinear regime, the photon bubble instability drives the collapse of the atmosphere starting from high altitudes. \citet{1997ApJ...478..663H} also find that the density structures that form in the instability are smaller with higher numerical resolution, as we also find.  However, they suggested that their photon bubbles tend to merge toward longer length scales in terms of their transport properties.  In particular, although the density fluctuations are on smaller length scales with increasing resolution, they found that the radiation energy density is spatially smoother because of radiation diffusion.  We find similar results but only at high altitude.  As shown in \autoref{fig:snapshot_800}, the spatial structure of radiation energy density closely tracks that of density even well into the nonlinear regime at low altitudes.  However, at high altitudes where the atmosphere is collapsing and radiation is able to diffuse much more quickly, we do see much broader structures in the radiation energy density, in agreement with \citet{1997ApJ...478..663H}.}

\section{Discussion and Conclusions}
\label{sec:discussion_conclusion}
All of the numerical experiments we have done in this paper are in preparation for a more global simulation of a neutron star accretion column. We have demonstrated here that we can successfully resolve photon bubbles and capture their nonlinear dynamics in a static medium.  In our next paper we will present our results on the impact of this instability on magnetically confined, accreting columns in Cartesian geometry.
While the simulations here had horizontally periodic boundary conditions and therefore lacked a boundary confined against radiation pressure by magnetic tension, we nevertheless successfully managed to constrain the gas motion by the strong magnetic fields in radiative RMHD simulations.  This is critical for a more global simulation in which the magnetic field will have to provide lateral confinement of the accretion column.  There radiation will likely escape the column mostly from the sides \citep{1976MNRAS.175..395B}, not the top as in the simulation here.  Whether the spatial resolution of the grid is as much of an issue in controlling the dynamics in that case, given the vertical shape of the photon bubble channels, remains to be seen.  A global simulation will also require accretion of material from the top boundary, which we have not yet incorporated here.  Such accretion is the only way that a steady-state column structure might develop against the photon bubble collapse that we found here. 

We summarize our conclusions as follows:

1. In the numerical simulation of the radiation-supported and magnetized atmosphere on neutron star, we resolve the characteristics of the photon bubble instability in the slow diffusion limit and explore the multi-mode behaviors depending on the radiative diffusion, propagation direction and wavelength.  In particular, modes grow most quickly at altitude where radiative diffusion is faster, with wave fronts that are significantly inclined to the vertical.  At depth where diffusion is more slow, the modes grow more slowly with wave fronts that are more aligned with the vertical magnetic field.
\newline\newline
2. We confirm the consistency between the numerical simulation and the linear theory of photon bubble instability. The simulation results illustrate the robustness of the current code framework of Athena++ in the linear phase and provide more insights into the non-linear dynamics caused by the photon bubble instability.  The faster growing, inclined modes at altitude spread downward.  However, the slow diffusion modes at depth eventually grow and result in vertical concentrations of density on magnetic field lines, separated by more tenuous reasons which allow for more rapid diffusion of photons.  This eventually always results in collapse of the atmosphere.  How this gets modified in the presence of additional mass supply from the top will be the subject of a future paper.
\newline\newline
3. We perform a resolution study and explore the resolution dependence of the photon bubble instability in the simulation, which suggests that the dynamical system involved with the photon bubble instability requires high resolution to capture the correct dynamical effects.  Low resolution simulations, while still collapsing, do so on longer time scales and with longer wavelengths because they are not able to resolve faster growing modes.  This resolution dependence persists well into the nonlinear regime, with the size scale of the nonlinear density structures scaling approximately with the grid cell size.  This contrasts with, other instabilities, e.g. Rayleigh-Taylor, whose linear growth rates increase toward shorter wavelength, but whose nonlinear structures have large length scale.  Here the nonlinear outcome of the photon bubble instability is dominated by the shortest resolved wavelengths \lz{until the viscous length scale is reached}.  This represents a numerical challenge for simulating this instability.
\newline\newline
The photon bubble instability causes all our simulations of static atmospheres to collapse, in agreement with the prediction of \citet{beg06}.  However, this still leaves open the question of what happens when fresh mass is supplied to an actual accreting column. Our work here lays the foundation for numerical simulations of magnetically confined, accreting columns on neutron stars, which enable us to resolve the photon bubble instability and study whether it affects existing models that assume a spatially smooth, stationary structure. We will publish the results of such simulations in our next paper. 

\section*{Acknowledgements}
\lz{We thank the referee for very useful comments that led to insights that greatly improved this paper.}  We thank Mitch Begelman, Matthew Middleton, Bryance Oyang, Jim Stone, and Chris White for useful conversations.  Chris White also provided invaluable help with the linear wave tests of the special relativistic radiation MHD module.  This work was supported in part by NASA Astrophysics Theory Program grant 80NSSC20K0525.
The simulations reported here were performed on computational facilities purchased with funds from the National Science Foundation (CNS-1725797) and administered by the Center for Scientific Computing (CSC). The CSC is supported by the California NanoSystems Institute and the Materials Research Science and Engineering Center (MRSEC; NSF DMR 1720256) at UC Santa Barbara. The Center for Computational Astrophysics at the Flatiron Institute is supported by the Simons Foundation. Resources supporting this work were also provided by the NASA High-End Computing (HEC) Program through the NASA Advanced Supercomputing (NAS) Division at Ames Research Center.

\section*{Data Availability}

All the simulation data reported here is available upon request to the authors.

\bibliographystyle{mnras}
\bibliography{paper1}

\appendix
\label{appendix}

\section{Derivation of photon bubble instability}
\label{sec:derivation_pbi}
\subsection{Conservation Laws in the Newtonian Limit}
\label{sec:pbi_setup_Newton}
A radiation-supported and magnetized plasma is unstable in a gravitational field \citep{1992ApJ...388..561A,1998MNRAS.297..929G}. Here we rederive the dispersion relation for the linear instability, \lz{incorporating the effects of radiation viscosity for the first time}.  The numerical simulations presented in this paper use special relativistic magnetohydrodynamics and a kinetic treatment of the radiation transfer.  However, for the linear instability analysis here, it is sufficient to use Newtonian equations with fluid restricted to move along the vertical ($\hat{z}$) magnetic field direction, and to treat the radiation transport within the diffusion approximation.  The system is then governed by the following equations:
\begin{align}
    & \frac{\partial\rho}{\partial t} + \frac{\partial}{\partial z}\left(\rho v\right) = 0
    \label{eq:mass_conserv_Newton}
    \quad,
    \\
    & \begin{multlined}[t]
    \rho\frac{\partial v}{\partial t} + \rho v\frac{\partial v}{\partial z} = -\frac{\partial P}{\partial z} - \rho g +\frac{\partial}{\partial x}\left(\eta\frac{\partial v}{\partial x}\right)
    \\
    +\frac{\partial}{\partial z}\left[\left(\frac{4\eta}{3}+\zeta\right)\frac{\partial v}{\partial z}\right]
    \quad,
    \end{multlined}
    \label{eq:mom_conserv_Newton}
    \\
    & \frac{\partial P}{\partial t} + v\frac{\partial P}{\partial z} + \frac{4}{3}P\frac{\partial v}{\partial z} = - \frac{1}{3}\frac{\partial F_x}{\partial x} - \frac{1}{3}\frac{\partial F_z}{\partial z}
    \label{eq:energy_conserv_Newton}
    \quad,
    \\
    & F_x = -\frac{c}{\rho\kappa}\frac{\partial P}{\partial x}
    \label{eq:rad_flux_horizontal}
    \quad, 
    \\
    & F_z = -\frac{c}{\rho\kappa}\frac{\partial P}{\partial z}
    \label{eq:rad_flux_vertical}
    \quad,
\end{align}
We approximate the total thermal pressure $P \simeq P_r$ as being entirely due to radiation. The horizontal radiation flux $F_x$ is perpendicular to the magnetic field and the vertical radiation flux $F_z$ is along the magnetic field. We also assume a constant vertical gravitational acceleration $g$, and assume a constant opacity $\kappa$ (dominated by Thomson scattering, assumed here to be isotropic). \lz{We have included shear ($\eta$) and bulk ($\zeta$) viscosity effects in the momentum equation (\autoref{eq:mom_conserv_Newton}), but have neglected viscous dissipation terms in the energy equation (\autoref{eq:energy_conserv_Newton}) as these would be nonlinear (second order) in the velocity.  \citet{1992ApJ...388..561A} also considered viscosity in their original analysis of the photon bubble instability, but only included the $4\eta/3$ term (cf. equation (4) from that paper).  They neglected the shear term arising from horizontal $(x)$ gradients in the vertical velocity $v$.  We find that these prove to be very important in the slow diffusion regime, where the most unstable modes have much larger horizontal than vertical gradients.}

In the static, plane-parallel equilibrium, mass conservation (\autoref{eq:mass_conserv_Newton}) and energy conservation (\autoref{eq:energy_conserv_Newton}) become trivial, leaving us with
\begin{align}
    & \frac{\partial P_0}{\partial z} = -\rho_0 g
    \label{eq:hb_mass_conserv_Newton}
    \quad,
    \\
    & F_{x0} = 0
    \quad, 
    \\
    & F_{z0} = \frac{cg}{\kappa}
    \quad,
    \label{eq:equilibrium}
\end{align}
where the subscript `0' refers to the equilibrium state. \newline
Linear perturbations about this equilibrium then evolve according to
\begin{align}
    & \frac{\partial\delta\rho}{\partial t} + \frac{\partial\rho_0}{\partial z}\delta v + \rho_0\frac{\partial\delta v}{\partial z} = 0
    \label{eq:lin_mass_conserv_Newton}
    \quad,
    \\
    & \begin{multlined}
    \rho_0\frac{\partial\delta v}{\partial t} = -\frac{\partial\delta P}{\partial z} - \delta\rho g +\frac{\partial}{\partial x}\left(\eta_0\frac{\partial\delta v}{\partial x}\right)
    \\
    +\frac{\partial}{\partial z}\left[\left(\frac{4\eta_0}{3}+\zeta_0\right)\frac{\partial\delta v}{\partial z}\right]
    \quad,
    \end{multlined}
    \label{eq:lin_mom_conserv_Newton}
    \\
    & \frac{\partial\delta P}{\partial t} + \delta v\frac{\partial P_0}{\partial z} +\frac{4}{3}P_0\frac{\partial\delta v}{\partial z} = - \frac{1}{3}\frac{\partial\delta F_x}{\partial x} - \frac{1}{3}\frac{\partial\delta F_z}{\partial z}
    \label{eq:lin_energy_conserv_Newton}
    \quad,
    \\
    & \delta F_x = -\frac{c}{\rho_0\kappa}\frac{\partial\delta P}{\partial x}
    \label{eq:lin_rad_flux_horizontal}
    \quad, 
    \\
    & \delta F_z = -\frac{c}{\rho_0\kappa}\left(\frac{\partial\delta P}{\partial z} + \delta\rho g\right) = \frac{c}{\kappa}\frac{\partial\delta v}{\partial t}
    \label{eq:lin_rad_flux_vertical}
    \quad,
\end{align}
where `$\delta$' in front of the variables refers to an Eulerian perturbation.

In what follows, we define some auxiliary parameters, most of which are consistent with the definitions used in \citet{1992ApJ...388..561A} and \citet{1998MNRAS.297..929G}. 
\begin{align}
    & c_r^2 = \frac{4P_0}{3\rho_0} \quad,
    && h = \frac{c_r^2}{g} \quad,
    \nonumber
    \\
    & N_0 = \frac{c_r}{h} \quad,
    && M_0 = \frac{c}{\rho_0\kappa h c_r} \quad,
    \label{eq:auxiliary_param}
    \\
    & k^2 = k_x^2 + k_z^2 \quad,
    && \mu = \frac{k_z}{k} \quad,
    \nonumber
\end{align}
where $c_r$ is the radiation sound speed and $h$ is the corresponding scale height. $N_0$ is the radiation sound crossing frequency, where $N_0^{-1}$ represents the time it takes to cross the scale height $h$ with the radiation sound speed $c_r$. $M_0$ is the Mach number of radiation diffusion, where $c/(\rho_0 \kappa h)$ is the radiation diffusion speed. We are in the slow diffusion regime if $M_0 \ll 1$. After the system is perturbed, we have the total wavenumber $k$, the horizontal wavenumber $k_x$ and the vertical wavenumber $k_z$, where $\mu$ is the cosine of the angle $\theta$ between the directions of wave propagation ($\hat{k}$) and magnetic field ($\hat{z}$). 

\subsection{Dispersion Relation}
\label{sec:pbi_dispersion_relation}

Because the equilibrium is static and homogeneous in the horizontal direction, we can, without loss of generality, assume that all perturbations depend on $x$ and time $t$ according to $\propto\exp[i(k_x x-\omega t)]$.  Equations (\ref{eq:lin_mass_conserv_Newton})-(\ref{eq:lin_rad_flux_vertical}) can then be combined to form two coupled ordinary differential equations in $z$:
\begin{align}
    \begin{split}
    -\omega^2\delta v = &\frac{i\omega}{\rho_0}\frac{d\delta P}{dz} + \frac{g}{\rho_0}\frac{d}{dz}(\rho_0\delta v)
    \\
    & -\frac{i\omega}{\rho_0}\frac{d}{dz}\left[\left(\frac{4\eta_0}{3}+\zeta_0\right)\frac{d\delta v}{dz}\right]+\frac{i\omega\eta_0}{\rho_0}k_x^2\delta v
    \quad,
    \end{split}
    \label{eq:dispersion_part1}
\end{align}
and
\begin{align}
    \begin{split}
    -i\omega\delta P = &\rho_0 g \delta v - \rho_0 c_r^2\frac{d\delta v}{dz}
    -\frac{k_x^2c}{3\kappa\rho_0}\delta P
    \\
    & -i\frac{c}{3\kappa\omega}\frac{d}{dz}\left[\frac{i\omega}{\rho_0}\frac{d\delta P}{dz}+\frac{g}{\rho_0}\frac{d}{dz}(\rho_0\delta v)\right]
    \quad.
    \end{split}
    \label{eq:dispersion_part2}
\end{align}
Apart from the neglect of nonlinearities, these equations are exact.  In the limit of infinite opacity, where radiative diffusion is negligible, they can be further combined to give an equation for vertical adiabatic sound waves in \lz{a viscous, strongly magnetized} inhomogeneous medium:
\begin{align}
    \begin{split}
    &\frac{d}{dz}\left(\rho_0c_r^2\frac{d\delta v}{dz}\right)+\rho_0\omega^2\delta v_z=
    \\
    &\mkern150mu i\omega\frac{d}{dz}\left[\left(\frac{4\eta_0}{3}+\zeta_0\right)\frac{dv}{dz}\right] - i\omega\eta k_x^2\delta v
    \quad.
    \end{split}
    \label{eq:adiabaticsoundwaves}
\end{align}
Note that the $(1/\rho_0)d(\rho_0\delta v)/dz$ term on the right hand side of (\autoref{eq:dispersion_part1}) completely cancels when this is done.

For finite opacity, equations (\ref{eq:dispersion_part1})-(\ref{eq:dispersion_part2}) cannot be combined directly, and we are forced to employ a short-wavelength vertical WKB approximation with perturbations having a $z$-dependence of the form $\exp\left( i\int^z k_z(z^\prime)dz^\prime \right)$.  We then directly replace all $z$-derivatives in equations (\ref{eq:dispersion_part1})-(\ref{eq:dispersion_part2}) with $i k_z$.  In so doing, we continue to maintain the cancellation of the $(1/\rho_0)d(\rho_0\delta v)/dz$ terms that led to (\autoref{eq:adiabaticsoundwaves}).  This treatment of the WKB approximation results in the following cubic dispersion relation
\begin{align}
    \begin{multlined}[t]
    \omega^3 + ik^2\left(\frac{c}{3\rho_0\kappa}+V\right)\omega^2 - \left(c_r^2\mu^2k^2+\frac{k^4cV}{3\rho_0\kappa}\right)\omega 
    \\
    - \frac{c}{3\rho_0\kappa}g\mu(1-\mu^2)k^3 = 0
    \quad,
    \end{multlined}
    \label{eq:dispersion_blaes}
\end{align}
\lz{where the viscous effects are in the quantity $V$ defined as}
\begin{equation}
    V\equiv\frac{1}{\rho_0}\left[\mu^2\left(\frac{4\eta_0}{3}+\zeta_0\right)+(1-\mu^2)\eta_0\right].
    \label{eq:V}
\end{equation}
\lz{If we neglect viscosity, this is almost the same as the cubic dispersion relation found by \citet{1992ApJ...388..561A}, the small differences arising from the difference in our WKB treatment.  It also}
recovers the slow diffusion dispersion relation (36) of \citet{1998MNRAS.297..929G} if we neglect the $\omega^3$ term. In the rapid diffusion regime as $M_0 \rightarrow \infty$ (small $\kappa$), (\autoref{eq:dispersion_blaes}) also recovers equation (34) of \citet{1998MNRAS.297..929G}.

As noted above, \lz{\citet{1992ApJ...388..561A} did actually consider radiation viscosity effects on vertical gradients in vertical velocity, and concluded that they would be unimportant for optically thick wavelengths.  For isotropic
Thomson scattering, the radiation shear viscosity is}
\begin{equation}
    \eta_0=\frac{8P_0}{9\kappa\rho_0c}
\end{equation}
\lz{\citep{MAS71}.  If we assume comparable bulk viscosity, then $V\sim c_r^2/(\kappa\rho c)$, in which case the viscous terms in the dispersion relation (\autoref{eq:dispersion_blaes}) are all negligible for optically thick wavelengths, in agreement with the assertion of \citet{1992ApJ...388..561A}.  However, this ignores the angle factor $\mu$.  In particular, the horizontal velocity gradients that were neglected in the viscous treatment of \citet{1992ApJ...388..561A} are important in the slow diffusion limit.  If we neglect the $\omega^3$ term and solve the resulting quadratic equation in the slow diffusion limit, we find that the instability growth rate peaks at wavenumber $k=2\pi/l_{\rm vis}$, where}
\begin{equation}
    l_{\rm vis}=\frac{4\pi}{3\kappa\rho}\frac{k_x}{k_z}
    =\frac{4\pi}{3\kappa\rho}\frac{\sqrt{1-\mu^2}}{\mu}.
    \label{eq:lvis}
\end{equation}
\lz{This can be much larger than the wavelength of unit optical depth $(\kappa\rho)^{-1}$ if
$k_x\gg k_z$, i.e. $\mu$ is small, and this is precisely the orientation of fastest growing slow diffusion modes.}

\begin{figure*}
	\includegraphics[width=\textwidth]{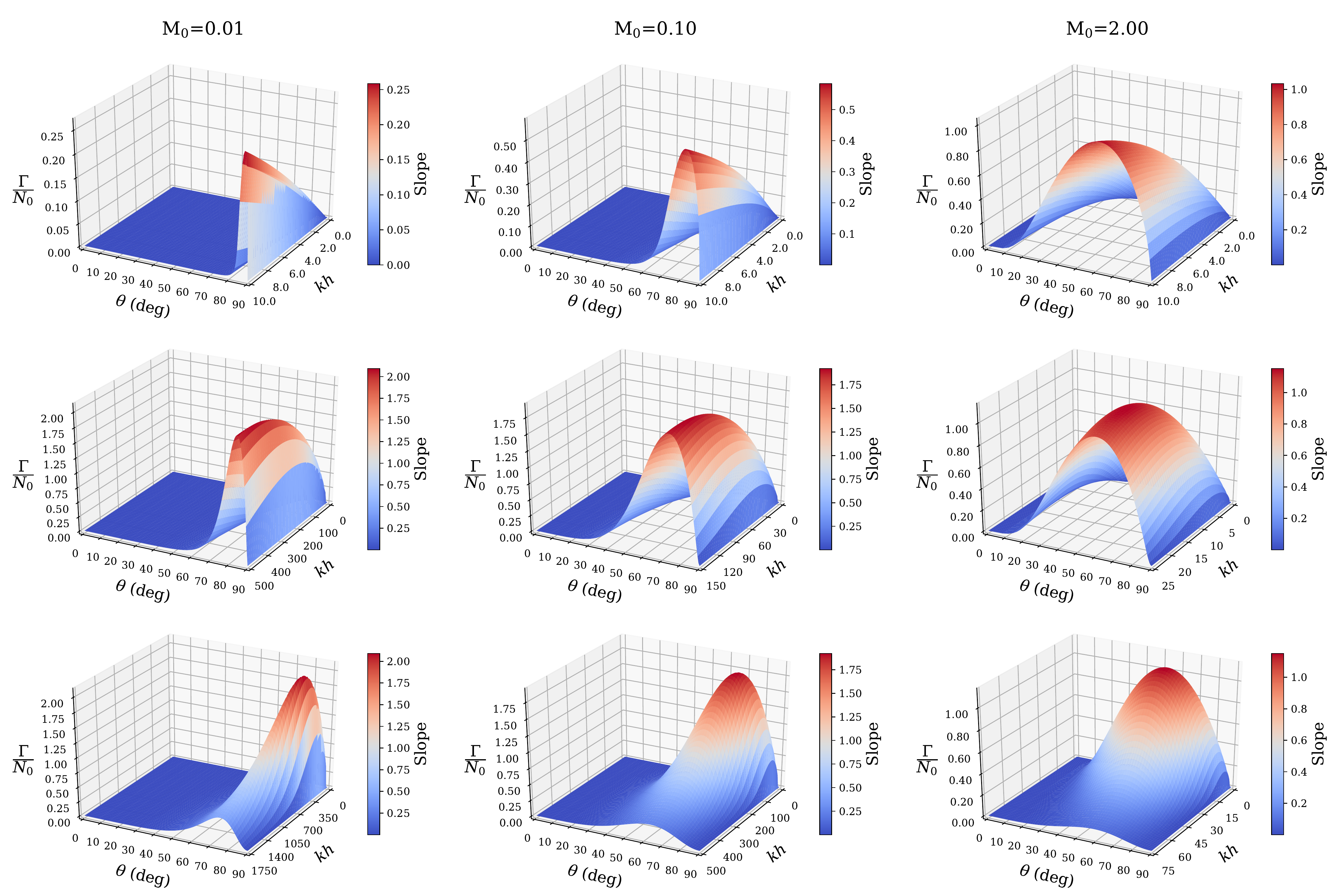}
    \caption{ Analytical solution of the instability growth rate based on the dispersion relation (\autoref{eq:dispersion_blaes_norm}),
    \lz{for $c_r/c=0.05$.}}
    \label{fig:growthRate3D}
\end{figure*}

For the convenience of numerical calculation, we can normalize (\autoref{eq:dispersion_blaes}) by the radiation sound crossing frequency $N_0$ as
\begin{align}
    \begin{split}
    &\left(\frac{\omega}{N_0}\right)^3 + i (kh)^2 \left(\frac{1}{3}M_0+\frac{V}{hc_r}\right) \left(\frac{\omega}{N_0}\right)^2
    \\
    &\mkern50mu -\left[\mu^2 (kh)^2 + \frac{1}{3} (kh)^4\left(\frac{V}{hc_r}\right)M_0\right] \left(\frac{\omega}{N_0}\right)
    \\
    &\mkern200mu -\frac{1}{3}M_0\mu(1-\mu^2)(kh)^3 = 0
    \quad. 
    \end{split}
    \label{eq:dispersion_blaes_norm}
\end{align}
The dispersion relation (\autoref{eq:dispersion_blaes_norm}) can be numerically solved in different diffusion regimes as a function of wavenumber and angle. The resulting instability growth rate $\Gamma=\mathrm{Im}\{\omega\}$ is depicted in \autoref{fig:growthRate3D} for various values of the diffusion parameter $M_0$. The peak of the instability growth rate shifts towards $90^{\circ}$ as radiation diffusion becomes slower (smaller $M_0$) and the instability grows faster as the wavelength becomes shorter, \lz{until the viscous scale $l_{\rm vis}$ is reached.} All of these characteristics can be found in our numerical simulation (see \hyperref[sec:results]{Results} in detail), which shows the consistency between the linear theory and the simulation. \lz{We stress that our simulation algorithm has no explicit radiation viscosity at all.  Instead, this comes for free because we are solving the angle-dependent radiative transfer equation.  Hence the agreement between the simulations and the analytic theory we have presented here is a nontrivial success.}

\subsection{Origin of Photon Bubble Instability}
\label{sec:origin_of_pbi}

\lz{We briefly discuss here the physical origin of the photon bubble instability in the slow diffusion regime, neglecting radiation viscosity whose effect is simply to damp the instability at small scales.}  A radiation pressure supported medium in a strong vertical magnetic field is subject to the spontaneous development of arbitrary fluctuations in the vertical distribution of density, while still maintaining hydrostatic equilibrium.  This is simply because the equilibrium equations (\ref{eq:hb_mass_conserv_Newton})-(\ref{eq:equilibrium}) admit any arbitrary vertical density distribution, which then sets the distribution of vertical radiation pressure gradient. As a result, the linearized equations of motion (\ref{eq:lin_mass_conserv_Newton})-(\ref{eq:lin_rad_flux_vertical}) admit an exact, zero-frequency static ($\delta v=0$) mode provided there are no horizontal variations in the perturbations $(k_x=0)$.  However, any horizontal variation in the perturbations, as much be present in a neutron star accretion column with finite horizontal width, will introduce a horizontal diffusive flux.  This in turn will cause some time-dependence, but if the diffusion is slow, vertical hydrostatic equilibrium can be maintained because the inertia term in the momentum equation will be small.  Let us first consider this case.

Neglecting the \lz{radiation viscosity and} inertia term in the momentum equation so that hydrostatic equilibrium is maintained in the perturbations, and applying the WKB approximation, the linearized equations (\ref{eq:lin_mass_conserv_Newton}), (\ref{eq:lin_mom_conserv_Newton}), (\ref{eq:lin_energy_conserv_Newton}), (\ref{eq:lin_rad_flux_horizontal}) and (\ref{eq:lin_rad_flux_vertical}) become
\begin{align}
    & -i\omega\delta\rho + \frac{\partial\rho_0}{\partial z}\delta v + \rho_0\frac{\partial\delta v}{\partial z} = 0
    \label{eq:lin_mass_conserv_Newton_static}
    \quad,
    \\
    & -i\omega\rho_0\delta v \simeq 0 = -\frac{\partial\delta P}{\partial z} - \delta\rho g
    \label{eq:lin_mom_conserv_Newton_static}
    \quad,
    \\
    & -i\omega\delta P + \delta v\frac{\partial P_0}{\partial z} + \frac{4}{3}P_0\frac{\partial\delta v}{\partial z} = - \frac{i}{3}k_x\delta F_x - \frac{1}{3}\frac{\partial\delta F_z}{\partial z}
    \label{eq:lin_energy_conserv_Newton_static}
    \quad,
    \\
    & \delta F_x = -\frac{c}{\rho_0\kappa}ik_x\delta P
    \label{eq:lin_rad_flux_horizontal_static}
    \quad, 
    \\
    & \delta F_z = -\frac{c}{\rho_0\kappa}\left(\frac{\partial\delta P}{\partial z} + \delta\rho g\right) = -i\frac{c}{\kappa}\omega\delta v \simeq 0
    \label{eq:lin_rad_flux_vertical_static}
    \quad. 
\end{align}
Note that the perturbed vertical flux naturally vanishes after neglecting the inertia term. We can also eliminate the perturbation in radiation
pressure using (\autoref{eq:lin_mass_conserv_Newton_static}) and (\autoref{eq:lin_mom_conserv_Newton_static})
\begin{equation}
    -i\omega\delta P + \delta v\frac{\partial P_0}{\partial z} = 0
    \label{eq:static_mode_eq1}
    \quad,
\end{equation}
So the energy equation (\autoref{eq:lin_energy_conserv_Newton_static}) becomes
\begin{equation}
    \frac{4}{3}P_0\frac{\partial\delta v}{\partial z} = - \frac{i}{3}k_x\delta F_x
    \label{eq:static_mode_eq2}
    \quad,
\end{equation}
indicating an equilibrium between adiabatic work and horizontal heat flow. 
We can then obtain a dispersion relation using equations (\ref{eq:hb_mass_conserv_Newton}), (\ref{eq:lin_rad_flux_horizontal_static}), (\ref{eq:static_mode_eq1}) and (\ref{eq:static_mode_eq2})
\begin{equation}
    4i \frac{P_0}{g}\frac{\partial}{\partial z}\left(\frac{\delta P}{\rho_0}\right)\omega = \frac{c}{\rho_0\kappa}\delta P k_x^2
    \label{eq:static_mode_dispersion}
    \quad,
\end{equation}
which indicates that the time-dependence, which is entirely oscillatory,
is determined by horizontal diffusion. To summarize, spontaneous density fluctuations can apparently be maintained in vertical hydrostatic equilibrium with slow oscillatory time-dependence driven by horizontal diffusion, provided fluid inertia is negligible.

However, including that small inertia actually drives this mode unstable.
If we include the inertia term and apply the WKB approximation in the $z$-direction, the perturbed quantities can be solved via equations (\ref{eq:lin_mass_conserv_Newton_static}), (\ref{eq:lin_mom_conserv_Newton_static}), (\ref{eq:lin_energy_conserv_Newton_static}), (\ref{eq:lin_rad_flux_horizontal_static}) and (\ref{eq:lin_rad_flux_vertical_static})
\begin{align}
    &\delta\rho = -\frac{1}{g}\frac{\partial\delta P}{\partial z} + \left(1-i\frac{\omega^2}{gk_z}-i\frac{1}{k_z}\frac{\partial\ln{\rho_0}}{\partial z}\right)^{-1}\frac{\omega^2}{g^2}\delta P
    \label{eq:drho_full}
    \quad,
    \\
    \begin{split}
    &\delta v = -\left(1-i\frac{1}{k_z}\frac{\partial\ln{\rho_0}}{\partial z}\right)^{-1}\frac{\omega}{\rho_0 g k_z}\frac{\partial\delta P}{\partial z}
    \\
    &\mkern+20mu +\left[\left(1-i\frac{1}{k_z}\frac{\partial\ln{\rho_0}}{\partial z}\right) \left(1-i\frac{\omega^2}{gk_z}-i\frac{1}{k_z}\frac{\partial\ln{\rho_0}}{\partial z}\right)\right]^{-1}\frac{\omega^3}{\rho_0 g^2 k_z}\delta P
    ,
    \end{split}
    \label{eq:dv_full}
    \\
    &\delta F_x = -\frac{c}{\kappa}\frac{k_x}{\rho_0 k_z}\frac{\partial\delta P}{\partial z}
    \label{eq:dFx_full}
    \quad,
    \\
    &\delta F_z = -\frac{c}{\rho_0\kappa}\left(\frac{\partial\delta P}{\partial z} + \delta\rho g\right) = -i\frac{c}{\kappa}\omega \delta v
    \label{eq:dFz_full}
    \quad.
\end{align}
We already know that the frequency without the inertia term is real (see \autoref{eq:static_mode_dispersion}). To explore the instability caused by the inertia term, we first need to define the frequency and perturbations of the mode without the inertia term
\begin{align}
    &\omega^{(0)} = -\left(1+i\frac{1}{k_z}\frac{\partial\ln{\rho_0}}{\partial z} \right)^{-1}\frac{cg}{4\kappa P_0}\frac{k_x^2}{k_z}
    \label{eq:omega_static}
    \quad,
    \\
    &\delta\rho^{(0)} = -\frac{1}{g}\frac{\partial\delta P}{\partial z}
    \label{eq:drho_static}
    \quad,
    \\
    &\delta v^{(0)} = -\left(1-i\frac{1}{k_z}\frac{\partial\ln{\rho_0}}{\partial z}\right)^{-1}\frac{\omega^{(0)}}{\rho_0 g k_z}\frac{\partial\delta P}{\partial z}
    \label{eq:dv_static}
    \quad,
    \\
    &\delta F_x^{(0)} = -\frac{c}{\kappa}\frac{k_x}{\rho_0 k_z}\frac{\partial\delta P}{\partial z}
    \label{eq:dFx_static}
    \quad,
    \\
    &\delta F_z^{(0)} = -\frac{c}{\rho_0\kappa}\left(\frac{\partial\delta P}{\partial z} + \delta\rho^{(0)} g\right)
    \label{eq:dFz_static}
    \quad, 
\end{align}
where the superscript $(0)$ refers to the mode without the inertia term. Next, we need to separate this mode from the full dispersion relation. Let us denote the superscript $(1)$ for the modifications introduced by the inertia term, where $f^{(1)}=f-f^{(0)}$. To simplify the calculation, we apply the following assumptions in advance: 

1. the short-wavelength approximation, which allows us to treat $\omega^2/(gk_z)$ and $\partial\ln{\rho_0}/(k_z\partial z)$ as small quantities. 

2. small frequency introduced by the inertia term (i.e. $\omega^{(1)} \ll \omega^{(0)}$). \newline
Thus, we can express the perturbed quantities in terms of $\delta P$ and keep the leading terms in equations (\ref{eq:drho_full})-(\ref{eq:drho_static}), (\ref{eq:dv_full})-(\ref{eq:dv_static}), (\ref{eq:dFx_full})-(\ref{eq:dFx_static}) and (\ref{eq:dFz_full})-(\ref{eq:dFz_static}) as follows
\begin{align}
    &\delta\rho^{(1)} \simeq \frac{\left(\omega^{(0)}\right)^2}{g^2}\delta P
    \label{eq:drho_inertia}
    \quad,
    \\
    &\delta v^{(1)} \simeq \frac{\omega^{(0)}}{\rho_0 g}\left[\frac{\left(\omega^{(0)}\right)^2}{gk_z}-i\frac{\omega^{(1)}}{\omega^{(0)}}\right]\delta P
    \label{eq:dv_inertia}
    \quad,
    \\
    &\delta F_x^{(1)} = 0
    \label{eq:dFx_inertia}
    \quad,
    \\
    &\delta F_z^{(1)} \simeq -\frac{c}{\rho_0\kappa g}\left(\omega^{(0)}\right)^2\delta P
    \label{eq:dFz_inertia}
    \quad. 
\end{align}
Specifically, we neglect orders higher than $O(\omega^2/(gk_z))$, $O(\partial\ln{\rho_0}/(k_z\partial z))$ and $O(\omega^{(1)}/\omega^{(0)})$ for $\delta\rho^{(1)}$ and $\delta F_z^{(1)}$, but keeping the leading terms of order $O(\omega^2/(gk_z))$ and $O(\omega^{(1)}/\omega^{(0)})$ for $\delta v^{(1)}$. The dispersion relation for $\omega^{(1)}$ can be determined by the corresponding linearized heat equation by subtracting the mode without the inertia term from (\autoref{eq:lin_energy_conserv_Newton_static})
\begin{equation}
    -i\omega^{(1)}\delta P - \rho_0 g \delta v^{(1)} + i\frac{4}{3}k_z P_0 \delta v^{(1)} = - \frac{1}{3}\frac{\partial\delta F_z^{(1)}}{\partial z}
    \quad. 
\end{equation}
We can eliminate $v^{(1)}$ and $\delta F_z^{(1)}$ by using (\autoref{eq:dv_inertia}) and (\autoref{eq:dFz_inertia}) to obtain the complex frequency introduced by the inertia term
\begin{equation}
    \omega^{(1)} = \frac{3\rho_0}{4P_0}\frac{1}{k_z^2}\left(\omega^{(0)}\right)^3 + i\left(\frac{c}{4\kappa P_0} - \frac{\omega^{(0)}}{gk_z}\right) \left(\omega^{(0)}\right)^2
    \quad.
\end{equation}
This can also be written in dimensionless form as
\begin{align}
    \frac{\omega^{(1)}}{N_0} &\simeq -\frac{1}{27}M_0^3\frac{(k_x h)^6}{(k_z h)^5} + i\frac{1}{27}M_0^3\frac{(kh)^2(k_x h)^4}{(k_z h)^4}
    \\
    &= -\frac{1}{27}M_0^3(kh)\frac{(1-\mu^2)^3}{\mu^5} + i\frac{1}{27}M_0^3(kh)^2\frac{(1-\mu^2)^2}{\mu^4}
    . 
\end{align}
\begin{figure}
    \centering
	\includegraphics[width=0.45\textwidth]{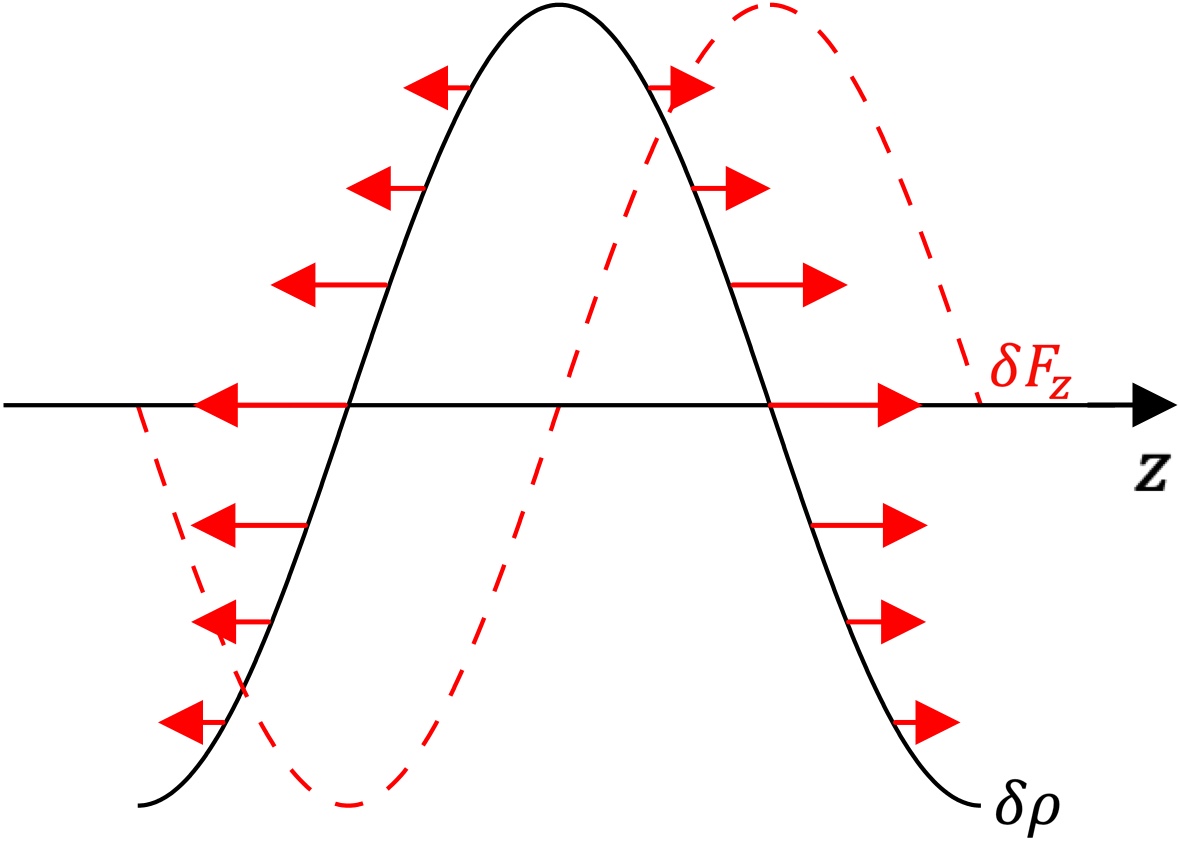}
    \caption{Phase differences of perturbations.}
    \label{fig:heat_flow}
\end{figure}
The growth rate of instability comes from the imaginary part of $\omega^{(1)}$, which is consistent with the approximation done by \citet{1992ApJ...388..561A} in his equation (41). The physical reasoning is well explained by \citet{1992ApJ...388..561A}. We follow his logic and first compare the phase differences of the perturbed quantities. By applying the short-wavelength approximation, we obtain the phase relation from (\autoref{eq:drho_full}) and (\autoref{eq:dFx_full})
\begin{align}
    \delta\rho &\propto -i\delta P
    \label{eq:phase_delay_drho}
    \quad,
    \\
    \delta F_z &\propto -\delta P
    \label{eq:phase_delay_dFz}
    \quad, 
\end{align}
which is shown in \autoref{fig:heat_flow}. Note that for spatially sinusoidal perturbations, $\delta\rho$ and $\delta P$ are $90^{\circ}$ out of phase. This is not an adiabatic perturbation, for which $\delta\rho$ and $\delta P$ would be in phase. Hence this mode is associated with nonzero entropy perturbations, and is in fact called an entropy mode. In this entropy mode, the perturbed radiation energy density and perturbed horizontal flux are dominated by the negligible inertia terms with the real frequency $\omega^{(0)}$. And the perturbed density is simply determined by the hydrostatic equilibrium. However, the inertia term is small but finite, which is sufficient to destabilize this mode. As shown in the \autoref{fig:heat_flow}, the perturbed vertical flux transfers photons from high-density regions to the low-density regions due to the $90^{\circ}$ phase delay. This tendency slowly evacuates the low-density regions with the radiation and leads to the increasing amplitude of perturbed density. \lz{Similar analysis can also be found in \citet{1992ApJ...388..561A}, where his figure~1 illustrates the same phase relation between the perturbed radiation flux and gas density that leads to the photon bubble instability. }

\section{Gravitation in weak field limit}
\label{sec:grav_source_terms}
Although we have included special relativistic MHD in our simulations in this paper, its effects are actually tiny and we could have done all the simulations using Newtonian physics.  The main reason we have incorporated special relativity is to prepare for doing simulations of actual magnetically confined, accreting columns of matter on neutron stars.  The primary relativistic effect there is that the 
Newtonian Alfv\'en speed in low density regions can exceed the speed of light, which would then require unreasonably small time steps in the simulation.  Even the free-fall speeds in the incoming accretion flow outside the accretion shock are only mildly relativistic.

However, gravity is important in the structure of the accretion column, as well as the dynamics of photon bubbles, and gravity is not defined in special relativity. One could in principle do full general relativistic MHD, but to do that with radiation would require incorporating the curved trajectories of photon geodesics in our transfer equation, and these effects are tiny over the short mean free paths within the optically thick portions of the column.  We therefore need only include weak field gravitational effects in our special relativistic MHD at lowest, essentially Newtonian, order.  We proceed here to reduce the full general relativistic conservation laws to the weak field limit and neglect all terms that are second order or higher in the corresponding Newtonian potential.

In isotropic coordinates, the weak field spacetime metric can be written as
\begin{equation}
    g_{\mu\nu} = \mathrm{diag}(-\mathbb{A}, \mathbb{B}, \mathbb{B}, \mathbb{B}) \quad,
    \label{eq:metric_iso}
\end{equation}
where $\mathbb{A}=\mathbb{A}(x^j)$ and $\mathbb{B}=\mathbb{B}(x^j)$ only vary in space. The corresponding Christoffel symbols are 
\begin{align}
    &\Gamma^{0}_{00} = 0
    &&\Gamma^{0}_{0j} = \Gamma^{0}_{j0} = \frac{1}{2\mathbb{A}} \partial_j \mathbb{A}
    &&\Gamma^{0}_{ij} = 0 \quad,
    \nonumber
    \\
    &\Gamma^{j}_{00} = \frac{1}{2\mathbb{B}} \partial_j \mathbb{A}
    &&\Gamma^{j}_{ij} = \Gamma^{j}_{ji} = \frac{1}{2\mathbb{B}} \partial_i \mathbb{B}
    &&\Gamma^{j}_{0i} = \Gamma^{j}_{i0} = 0 \quad,
    \label{eq:christoffel_iso}
    \\
    &
    &&\Gamma^{j}_{ii}\big{|}_{i\ne j} = -\frac{1}{2\mathbb{B}} \partial_j \mathbb{B}
    &&\Gamma^{j}_{ik}\big{|}_{i\ne j\ne k} = 0 \quad.
    \nonumber
\end{align}
The normalization of the four-velocity implies
\begin{equation}
    u^0 u^0 = \frac{1}{\mathbb{A}-\mathbb{B}v^2} \quad. 
\end{equation}
We now consider the conservation laws of particle number, momentum and energy:
\begin{align}
    \nabla_{\mu} (\rho u^{\mu}) &= 0 \quad, \label{eq:particle_conserv_gr}
    \\
    \nabla_{\mu} T^{\mu\nu} &= 0 \quad, \label{eq:gas_conserv_gr}
\end{align}
where the stress-energy tensor $T^{\mu\nu} = w_g u^{\mu} u^{\nu} + P_g g^{\mu\nu}$, gas enthalpy $w_g = \rho + \dfrac{\gamma}{\gamma-1}P_g$ and $\gamma$ is the gas adiabatic index. For simplicity, we neglect radiation and magnetic fields as that is all that is necessary to derive the form of the gravitational source terms.

Expanding (\autoref{eq:particle_conserv_gr}) and (\autoref{eq:gas_conserv_gr}) in the metric~(\ref{eq:metric_iso})
\begin{align}
    \partial_0 (\rho u^{0}) + \partial_j (\rho u^{j}) = &-\frac{1}{2}\left(\frac{1}{\mathbb{A}}\partial_j\mathbb{A} + \frac{3}{\mathbb{B}}\partial_j\mathbb{B}\right) \rho u^j
    \quad, \label{eq:particle_conserv_iso}
    \\
    \begin{split}
    \partial_0 T^{0i} + \partial_jT^{ij} = {}&
    -\left(\frac{1}{2\mathbb{A}}\partial_j\mathbb{A} + \frac{2}{\mathbb{B}}\partial_j\mathbb{B}\right)T^{ij} 
    \\
    & -\frac{1}{2\mathbb{B}}T^{ik}\partial_k\mathbb{B}\bigg{|}_{k\ne i} -\frac{1}{2\mathbb{B}} T^{00}\partial_i\mathbb{A}
    \\
    & +\frac{1}{2\mathbb{B}} T^{jj}\partial_i\mathbb{B} \bigg{|}_{j\ne i}
    \quad,
    \end{split}
    \label{eq:mom_conserv_iso}
    \\
    \begin{split}
    \partial_0 T^{00} + \partial_j T^{j0} =& -\left(\frac{1}{\mathbb{A}}\partial_j\mathbb{A} + \frac{3}{2\mathbb{B}}\partial_j\mathbb{B}\right)T^{j0} 
    \\
    & -\frac{1}{2\mathbb{A}}T^{0j}\partial_j\mathbb{A}
    \quad. 
    \end{split}
    \label{eq:energy_conserv_iso}
\end{align}
In the weak field limit, the metric~(\ref{eq:metric_iso}) can be specified as
\begin{equation}
    \begin{cases}
    \mathbb{A} = 1+2\phi
    \\
    \mathbb{B} = 1-2\phi
    \end{cases}
    ,\textrm{ where the potential } \phi=-\frac{GM}{c^2r} \textrm{ is a small quantity.} 
\end{equation}
Then, we expand equations~(\ref{eq:particle_conserv_iso}), (\ref{eq:mom_conserv_iso}) and (\ref{eq:energy_conserv_iso}) to first order of $\phi$                            \begin{align}
    & \partial_0 (\rho\Gamma) + \partial_j (\rho\Gamma v^{j}) = S_{\mathrm{gr}1}
    \quad,
    \label{eq:particle_conserv_approx}
	\\
	& \partial_0(w_g\Gamma^2v^i) + \partial_j(w_g\Gamma^2 v^i v^j + P_g\delta^{ij}) = S_{\mathrm{gr}2}^i
    \quad,
    \label{eq:mom_conserv_approx}
	\\
	& \partial_0\left( w_g\Gamma^2 - P_g \right) + \partial_j (w_g\Gamma^2 v^j) = S_{\mathrm{gr}3}
    \quad,
    \label{eq:energy_conserv_approx}
\end{align}
where the terms related to the gravitational field are 
\begin{align}
    S_{\mathrm{gr}1}= &4\Gamma^2(\partial_0\Gamma + v^j\partial_j\Gamma)\rho\phi + \Gamma(2\Gamma^2+1)\rho v^j\partial_j\phi
    \quad,
    \label{eq:Sgr1_approx}
    \\
    S_{\mathrm{gr}2}^i=&\begin{multlined}[t]
    8w_g\Gamma^3\phi v^i(\partial_0\Gamma + v^j\partial_j\Gamma) - 4\phi\Gamma^2\partial_iP_g 
    \\
    +2\Gamma^2(2\Gamma^2+1)w_gv^iv^j\partial_j\phi - (2\Gamma^2-1)w_g\partial_i\phi
    \quad,
    \end{multlined}
    \label{eq:Sgr2_approx}
    \\
    S_{\mathrm{gr}3}=&\begin{multlined}[t]
    8w_g\Gamma^3\phi(\partial_0\Gamma + v^j\partial_j\Gamma) + 4(\Gamma^2-1)\phi\partial_0P_g 
    \\
    +2\Gamma^2(2\Gamma^2-1)w_gv^j\partial_j\phi
    \quad.
    \label{eq:Sgr3_approx}
    \end{multlined}
\end{align}
We can check the consistency with the Newtonian limit by
applying approximations on (\autoref{eq:particle_conserv_approx}), (\autoref{eq:mom_conserv_approx}) and (\autoref{eq:energy_conserv_approx}) in different orders of small quantities. Keeping the first order of $v$ and $\phi$, we recover the Newtonian continuity equation  
\begin{align}
    &\partial_0\rho + \partial_j(\rho v^{j}) = 0
    \quad.
\end{align}
Define the isothermal sound speed as $c_0^2=P_g/\rho$. Neglecting the orders higher than $O(c_0^2)$, $O(v^2)$ and $O(\phi)$, we recover the Newtonian momentum conservation
\begin{equation}
    \partial_0(\rho v^i) + \partial_j(\rho v^iv^j) = -\partial_i P_g - \rho\partial_i\phi
    \quad.
\end{equation}
Neglecting the orders higher than $O(s^3)$, where $c_0\sim v\sim O(s)$ and $\phi\sim O(s^2)$, we recover the Newtonian energy conservation
\begin{equation}
    \partial_0\left(\frac{1}{\gamma-1}P_g + \frac{1}{2}\rho v^2 + \rho\phi\right) + \partial_j\left[\left(\frac{\gamma}{\gamma-1} P_g + \frac{1}{2}\rho v^2 + \rho\phi\right)v^j\right] = 0
    \quad.
\end{equation}

Therefore, in order to capture gravitational effects to lowest (Newtonian) order,
we can apply the following source terms in the framework of Athena++ RMHD module to mock up the gravitational effect near the neutron star surface.
\begin{align}
    S_{\mathrm{gr}1} \simeq& \Gamma(2\Gamma^2+1)\rho v^j\partial_j\phi
    \quad,
    \\
    S_{\mathrm{gr}2}^i \simeq& +2\Gamma^2(2\Gamma^2+1)w_gv^iv^j\partial_j\phi - (2\Gamma^2-1)w_g\partial_i\phi
    \quad,
    \\
    S_{\mathrm{gr}3} \simeq& 2\Gamma^2(2\Gamma^2-1)w_gv^j\partial_j\phi
    \quad.
\end{align}

\section{Athena++ code modification}
\label{sec:code_modify}
Special relativity is essential if we want to maintain finite Alfv\'en speeds and reasonable time steps when the magnetic field is strong in low density regions. Hence, we need to adjust the radiation module \citep[][]{2014ApJS..213....7J} to couple with RMHD \citep[][]{2011ApJS..193....6B} in Athena++. In this section, we give an overview of the numerical treatment for solving the radiation and discuss our modifications based on the original framework. The radiation module solves the frequency-integrated radiative transfer equation 
\begin{equation}
    \partial_0I(n^i) + n^j\partial_jI(n^i) = S_r
    \quad, 
\end{equation}
where the source term $S_r = \eta(n^i) - \chi(n^i) I(n^i)$. 
The radiation source term and advective term are separately used to update the intensity by using an operator split approach.  The source term updates the intensity in the fluid rest frame: 
\begin{equation}
    \partial_0\bar{I} = \Gamma(1-v_jn^j)\left(\bar{\eta} - \bar{\chi} \bar{I} \right)
    \quad, 
\end{equation}
then the advection term updates the intensity in the lab frame:
\begin{equation}
    \partial_0I + n^j\partial_jI(n^i) = 0
    \quad.
\end{equation}
Since the radiation source term is defined in the co-moving frame, we modify the code so that the RMHD primitive variables are updated in time before being used in the radiation module. The radiation source term incorporates elastic scattering, absorption and Compton scattering (\citealt{2003ApJ...596..509B}; \citealt{2009ApJ...691...16H})
\begin{align}
    \begin{split}
    \partial_0\bar{I} = \Gamma(1-v_jn^j)\bigg[ &\rho\kappa_s(\bar{J}-\bar{I})
    \\
    +&\rho\kappa_R\left(\frac{a_r T_g^4}{4\pi}-\bar{I}\right) +\rho(\kappa_P-\kappa_R)\left(\frac{a_r T_g^4}{4\pi}-\bar{J}\right)
    \\
    +&\rho\kappa_s\frac{4(T_g-\bar{T}_r)}{T_e}\bar{J} \bigg]
    \quad.
    \end{split}
    \label{eq:rad_source_term_equation}
\end{align}
Equation~(\ref{eq:rad_source_term_equation}) needs to be solved together with the gas temperature, which can be determined from the energy exchange between gas and radiation in the fluid frame
\begin{equation}
    \frac{\rho R}{\gamma-1}\partial_{\bar{0}}T_g = -\rho\kappa_R(a_rT_g^4 - \bar{E}_r) -\rho\kappa_s\frac{4(T_g-\bar{T}_r)}{T_e}\bar{E}_r
    \quad.
    \label{eq:gas_rad_energy_exchange}
\end{equation}
Note that the code framework integrates over lab frame time $t$ rather than fluid frame time $\bar{t}$. The original code treats $\partial_0=\partial_{\bar{0}}$ in Newtonian physics. However, we need to distinguish them in special relativity. Recall the Lorentz transformation of the time derivative
\begin{equation}
    \partial_{\bar{0}} = \Gamma(\partial_0+v^{j}\partial_j)
    \quad.
\end{equation}
If we simply plug this into (\autoref{eq:gas_rad_energy_exchange}), we find that the derivatives of space and time are mixed again, which breaks the operator splitting approach. Thus, we apply the approximation $\partial_{\bar{0}} \simeq \Gamma\partial_0$, which will be valid if $ \Delta t < \mathrm{min}\left(\frac{\Delta x^j}{|v^j|}\right)$. This criterion should be easily satisfied unless the regime is ultra-relativistic, which then requires the CFL number $C_{\mathrm{CFL}}$ to be small enough
\begin{equation}
    C_{\mathrm{CFL}} < \frac{ \mathrm{min}\left( \Delta x^j/|v^j| \right) }{ \mathrm{min}\left( \Delta x^j/|\lambda^j| \right) }
    \quad, 
    \label{eq:criterion_cfl}
\end{equation}
where $\lambda^j$ is the fastest wavespeed and $C_{\mathrm{CFL}}$ is a positive constant smaller than unity. We are interested in regimes near the surface of a neutron star, where the magnetic field is strong. For the static column, the velocity is $\sim 0$. For the accreting column, the free-fall speed can reach up to $\sim 0.64c$. The fastest wavespeed is $\sim c$, which is determined by the Alfv\'en wave in the low density regions. Therefore, the criterion~(\autoref{eq:criterion_cfl}) is easily satisfied in both cases. 

The regime we work in has a strong magnetic field and negligible gas pressure. The magnetic pressure can exceed gas pressure by factor of $10^{10}$, which makes the gas temperature hard to resolve. Although the dynamics of the system is mainly driven by radiation pressure and magnetic confinement, the gas temperature is still important in the calculation of the radiation source term. Hence, we update the primitive variable inversion algorithm as described in \citet{doi:10.1137/140956749} in the original framework of RMHD for more robust numerical behavior.

With the above modifications, we checked that the code passes the thermal equilibrium test and momentum conservation test following \citealt{2014ApJS..213....7J}. We also performed linear wave tests in the diffusion limit and the tests converged to greater than first order in spatial resolution.

\bsp	
\label{lastpage}

\end{CJK*}
\end{document}